\documentclass{emulateapj}

\usepackage{amsfonts}
\usepackage{amsmath}
\usepackage{graphicx}

\def\so{1}
\def\sotxt{Steward Observatory, University of Arizona, 933 N. Cherry Avenue, Tucson, AZ 85721, USA; rskibba@as.arizona.edu}
\def\wyoming{2}
\def\wyomingtxt{Department of Physics \& Astronomy, University of Wyoming, Laramie, WY 82071, USA}
\def\mpia{3}
\def\mpiatxt{Max-Planck-Institut f\"{u}r Astronomie, K\"{o}nigstuhl 17, 69117 Heidelberg, Germany}
\def\dark{4}
\def\darktxt{Dark Cosmology Centre, Niels Bohr Institute, University of Copenhagen, Juliane Maries Vej 30, 2100 Copenhagen \O, Denmark}
\def\umass{5}
\def\umasstxt{Department of Astrophysics, University of Massachusetts, 710 North Pleasant Street, Amherst, MA, 01003, USA}
\def\leiden{6}
\def\leidentxt{Sterrewacht Leiden, Leiden University, P.O. Box 9513, 2300 RA Leiden, The Netherlands}
\def\arcetri{7}
\def\arcetritxt{INAF-Osservatorio Astrofisico di Arcetri, Largo Enrico Fermi 5, 50125 Firenze, Italy}
\def\cambridge{8}
\def\cambridgetxt{Institute of Astronomy, University of Cambridge, Madingley Road, Cambridge CB3 0HA, UK}
\def\iap{9}
\def\iaptxt{Institut d'Astrophysique de Paris, CNRS, Universit\'{e} Pierre \& Marie Curie, UMR 7095, 98bis bd Arago, 75014 Paris, France}
\def\ssc{10}
\def\ssctxt{Spitzer Science Center, California Institute of Technology, MC 314-6, Pasadena, CA 91125, USA}
\def\maryland{11}
\def\marylandtxt{Department of Astronomy, University of Maryland, College Park, MD 20742, USA}
\def\toledo{12}
\def\toledotxt{Department of Physics and Astronomy, University of Toledo, Toledo, OH 43606, USA}
\def\stsci{13}
\def\stscitxt{Space Telescope Science Institute, 3700 San Martin Drive, Baltimore, MD 21218, USA}
\def\ipac{14}
\def\ipactxt{NASA Herschel Science Center, IPAC, California Institute of Technology, Pasadena, CA 91125, USA}
\def\suny{15}
\def\sunytxt{Department of Physics and Astronomy, SUNY Stony Brook, Stony Brook, NY 11794-3800, USA}
\def\cea{16}
\def\ceatxt{CEA/DSM/DAPNIA/Service d'Astrophysique, UMR AIM, CE Saclay, 91191 Gif sur Yvette Cedex, France}
\def\mcmaster{17}
\def\mcmastertxt{Department of Physics \& Astronomy, McMaster University, Hamilton, Ontario L8S 4M1, Canada}

\slugcomment{}

\shorttitle{Dust and Stellar Emission of Galaxies}
\shortauthors{R. A. Skibba, C. W. Engelbracht, et al.}
\journalinfo{Accepted to the Astrophysical Journal on xx xx xxxx}

\begin{document}

\title{The Emission by Dust and Stars of Nearby Galaxies in the Herschel KINGFISH Survey}

\author
 {
  Ramin~A.~Skibba\altaffilmark{\so}, 
  Charles~W.~Engelbracht\altaffilmark{\so}, 
  Daniel~Dale\altaffilmark{\wyoming}, 
  Joannah~Hinz\altaffilmark{\so}, 
  Stefano~Zibetti\altaffilmark{\mpia,\dark}, 
  Alison~Crocker\altaffilmark{\umass}, 
  Brent~Groves\altaffilmark{\leiden,\mpia}, 
  Leslie~Hunt\altaffilmark{\arcetri}, 
  Benjamin~D.~Johnson\altaffilmark{\cambridge,\iap}, 
  Sharon~Meidt\altaffilmark{\mpia}, 
  Eric~J.~Murphy\altaffilmark{\ssc}, 
  %
  Philip~Appleton\altaffilmark{\ipac},
  Lee~Armus\altaffilmark{\ssc}, 
  Alberto~Bolatto\altaffilmark{\maryland}, 
  Bernhard~Brandl\altaffilmark{\leiden}, 
  Daniela~Calzetti\altaffilmark{\umass},
    Kevin~Croxall\altaffilmark{\toledo}, 
    Maud~Galametz\altaffilmark{\cambridge}, 
  Karl~D.~Gordon\altaffilmark{\stsci}, 
  Robert~C.~Kennicutt\altaffilmark{\cambridge},
  Jin~Koda\altaffilmark{\suny}, 
  Oliver~Krause\altaffilmark{\mpia}, 
  Edward~Montiel\altaffilmark{\so},
  Hans-Walter~Rix\altaffilmark{\mpia}, 
  H\'{e}l\`{e}ne~Roussel\altaffilmark{\iap}, 
    Karin~Sandstrom\altaffilmark{\mpia}, 
  Marc~Sauvage\altaffilmark{\cea}, 
  Eva~Schinnerer\altaffilmark{\mpia}, 
  J.D.~Smith\altaffilmark{\toledo}, 
  Fabian~Walter\altaffilmark{\mpia}, 
  Christine~D.~Wilson\altaffilmark{\mcmaster}, 
  Mark~Wolfire\altaffilmark{\maryland} 
 }

\altaffiltext{\so}{\sotxt}
\altaffiltext{\wyoming}{\wyomingtxt}
\altaffiltext{\mpia}{\mpiatxt}
\altaffiltext{\dark}{\darktxt}
\altaffiltext{\umass}{\umasstxt}
\altaffiltext{\leiden}{\leidentxt}
\altaffiltext{\arcetri}{\arcetritxt}
\altaffiltext{\cambridge}{\cambridgetxt}
\altaffiltext{\iap}{\iaptxt}
\altaffiltext{\ssc}{\ssctxt}
\altaffiltext{\maryland}{\marylandtxt}
\altaffiltext{\stsci}{\stscitxt}
\altaffiltext{\ipac}{\ipactxt}
\altaffiltext{\suny}{\sunytxt}
\altaffiltext{\cea}{\ceatxt}
\altaffiltext{\toledo}{\toledotxt}
\altaffiltext{\mcmaster}{\mcmastertxt}






\begin{abstract}

Using new far-infrared imaging from the \textit{Herschel} Space Observatory with ancillary data 
from ultraviolet (UV) to submillimeter (submm) wavelengths, 
we estimate the total emission from dust and stars of 62 
nearby galaxies in the KINGFISH survey in a way that is as empirical and 
model-independent as possible.
We collect and exploit these data 
in order to measure from the spectral energy distributions (SEDs)  
precisely how much stellar radiation is intercepted and 
re-radiated by dust, and how this quantity varies with galaxy properties.
By including SPIRE data, we are more sensitive to emission from cold dust grains than previous analyses 
at shorter wavelengths, allowing for more accurate estimates of dust temperatures and masses. 

The dust/stellar flux ratio, which we measure by integrating the SEDs, 
has a range of nearly three decades (from $10^{-2.2}$ to $10^{0.5}$). 
The inclusion of SPIRE data shows that estimates based on data not reaching 
these far-IR wavelengths are biased low by $17\%$ on average. 
We find that the dust/stellar flux ratio varies with morphology 
and total infrared (IR) luminosity, with dwarf galaxies having faint luminosities, spirals having 
relatively high dust/stellar ratios and IR luminosities, and some early-types having low dust/stellar ratios. 
We also find that dust/stellar \textit{flux} ratios are related to gas-phase metallicity 
($\overline{\mathrm{log}(f_\mathrm{dust}/f_\ast)}=-0.66\pm0.08$ and $-0.22\pm0.12$ 
for metal-poor and intermediate-metallicity galaxies, respectively), 
while the dust/stellar \textit{mass} ratios are less so (differing by $\approx0.2$~dex); 
the more metal-rich galaxies span a much wider range of the flux ratios. 
In addition, the substantial scatter between dust/stellar flux and dust/stellar mass 
indicates that the former is a poor proxy of the latter. 
Comparing the dust/stellar flux ratios and dust temperatures, we also 
show that early-types tend to have slightly warmer temperatures (by up to 5~K) 
than spiral galaxies, which may be due to more intense interstellar radiation fields,  
or possibly to different dust grain compositions. 
Finally, we show that 
early-types and early-type spirals have a strong correlation 
between the dust/stellar flux ratio and  
specific star formation rate, which suggests that the relatively bright 
far-IR emission of some of these galaxies is due to ongoing (if limited) 
star formation as well as to the radiation field from older stars, which is 
heating the dust grains. 
\end{abstract}


%

\section{Introduction}

Nearly half of the bolometric luminosity of the Universe is channeled through the mid- and far-infrared (IR) emission of galaxies (e.g., Hauser \& Dwek 2001). 
This spectral region also probes important physical properties of galaxies, such 
as their metal content, dust content, and cold gas content (e.g., Draine et al.\ 2007). 
To understand galaxies we must first understand the physical processes that regulate their evolution,  
including the formation of stars and the interstellar radiation field (ISRF), and 
the return of radiant energy from these stars into the interstellar medium (ISM). 
\textit{Herschel} stands poised to make major breakthroughs in these areas, 
by mapping galaxies in the far-IR with unprecedented spatial resolution.

Here we perform an empirical analysis of the galaxies in the KINGFISH project (Key Insights on Nearby Galaxies: a Far-Infrared Survey with \textit{Herschel}; P.I.: R.~C. Kennicutt), an imaging and spectroscopic survey of 61 nearby ($d\,<\,30\,\mathrm{Mpc}$) galaxies, chosen to cover the full range of integrated properties and local ISM environments found in the nearby Universe.
KINGFISH is closely following the observing strategy of SINGS (\textit{Spitzer} Infrared Nearby Galaxies Survey; Kennicutt et al.\ 2003), by mapping galaxies in their entirety with PACS (Photodetector Array Camera and Spectrometer; Poglitsch et al.\ 2010) at 70, 100, and 160$\mu$m and SPIRE (Spectral and Photometric Imaging REceiver; Griffin et al.\ 2010) at 250, 350, and 500$\mu$m. 

The physical processes contributing to the energetic output of a galaxy can be probed 
by constructing its spectral energy distribution (SED).  
Most of the stellar light is emitted in the ultraviolet (UV) 
to near-IR domain, with the short-lived, massive stars dominating the UV and the more numerous older stars the near-IR. 
Dust, produced by the aggregations of metals injected into the ISM by massive 
stars through stellar winds and supernovae, absorbs the stellar light and re-emits it 
in the IR and submillimeter (submm) domains. 

Our goal is to determine empirically how much starlight escapes galaxies and how 
much is intercepted by dust, as a function of other galaxy properties. 
In particular, we measure how the ratio of dust/stellar flux is correlated with 
properties such as morphology, metallicity, total IR luminosity, dust/stellar mass, 
dust temperature, and star formation rate. 
The KINGFISH sample is ideal for such an analysis, because these nearby galaxies have been 
extensively studied with a variety of telescopes and models. 
Many of the galaxies' properties are already well-determined, and others are now 
better constrained with data from \textit{Herschel}. 
A secondary goal is to determine which subset (if any) of galaxies observed by KINGFISH 
could be plausible local counterparts of galaxies dominating the extragalactic background light (EBL).

While our approach is an empirical one, there are alternative theoretical 
approaches as well.  A variety of different models have been applied to galaxy SEDs
(e.g., Fioc \& Rocca-Volmerange 1997; Silva et al.\ 1998; Devriendt et al.\ 1999; 
see review by Walcher et al.\ 2011), and there 
has been much recent work attempting to model both the stellar and dust SEDs 
of galaxies, over a wide range of wavelengths (e.g., Johnson et al.\ 2007; 
da Cunha et al.\ 2008; Noll et al.\ 2009; Buat et al.\ 2011; Popescu et al.\ 2011). 
Some of the results of these studies can be compared to ours (see Section~\ref{results}). 

This paper is organized as follows. 
In the next section, we describe the KINGFISH sample, and how we measure the galaxies' 
flux densities, from UV to submm wavelengths. 
We explain how we integrate the SEDs to obtain dust/stellar 
flux ratios in Section~\ref{dusttostellar}. 
Then in Section~\ref{results}, we present our results, showing how the dust/stellar 
flux ratios are correlated with various galaxy properties.  
In Section~\ref{EBL}, we discuss how some galaxies in the KINGFISH sample could be considered 
local counterparts of galaxies contributing to the EBL. 
We end with a summary of our results and conclusions in Section~\ref{discussion}.

\section{Data}


Our sample consists of 62 nearby galaxies, of which 61 are in KINGFISH (Kennicutt, Calzetti et al.\, in prep.), 
and the other is M~33, observed by the \textit{Herschel} M~33 Extended Survey (HERM33ES; Kramer et al.\ 2010). 
The sample is chosen to cover a large range of galaxy properties, 
such as morphological type, luminosity, metallicity, star formation rate, surface brightness, extinction, gas mass, dust content, radiation field strength, and ISM environment (see Kennicutt et al.\ 2003); however, the sample is not complete with respect to these properties.

\subsection{Flux Densities}

18 galaxies in our sample are in the \textit{Spitzer} Local Volume Legacy (LVL) survey\footnote{\texttt{http://www.ast.cam.ac.uk/IoA/research/lvls}}, and for these we use the global flux densities measured by Dale et al.\ (2009). 
57 of the galaxies are in SINGS, and for those that are not in LVL, we use the 
flux densities measured by Dale et al.\ (2007), 
or when applicable, the updated values presented by Mu\~{n}oz-Mateos et al.\ (2009b).

We obtained UV data from the Galaxy Evolution Explorer (GALEX; Martin et al.\ 2005; 
1528 and 2271\AA$\,$ wavelengths); 
optical data from either the Sloan Digital Sky Survey (SDSS; York et al.\ 2000; $ugriz$ bands) 
or Kitt Peak ($BVRI$ bands); 
near-IR data from the Two Micron All Sky Survey (2MASS; Skrutskie et al.\ 2006; $JHK$ bands); 
mid- and far-IR data from \textit{Spitzer}'s Infrared Array Camera (IRAC; Fazio et al.\ 2004; 3.6, 4.5, 5.8, 8 $\mu$m) 
and Multiband Imaging Photometer (MIPS; Rieke et al.\ 2004; 24, 70, 160 $\mu$m), 
and 100 $\mu$m from the Infrared Astronomical Satellite (IRAS; Soifer et al.\ 1987), when available; 
and submm data for one third of the galaxies from the Submillimeter Common-User Bolometric Array (SCUBA; Holland et al.\ 1999; 450 and 850 $\mu$m). 

For the two galaxies that were neither in SINGS nor LVL, IC~342 and NGC~2146, we obtained flux densities from various sources. 
We obtained UV magnitudes from GALEX (using the values quoted in Gil de Paz et al.\ 2007 for NGC~2146), and converted them to fluxes using the calibration in Morrissey et al.\ (2007). 
We obtained optical fluxes from Buta \& McCall (1999) and Marcum et al.\ (2001), respectively. 
We obtained 2MASS fluxes from Jarrett et al.\ (2003), using the calibration in Cohen et al.\ (2003). 
We measured Spitzer fluxes from their images, 
and we add the IRAS 100$\mu$m flux for NGC~2146. 
Lastly, M~33 (NGC~598) lacked optical fluxes in LVL, so we computed them from the data in Massey et al.\ (2006).
%

To these data we add \textit{Herschel} far-infrared data from SPIRE (250, 350, and 500 $\mu$m). 
We have obtained SPIRE observations of 61 KINGFISH galaxies, 
including six galaxies (NGC~4254, NGC~4321, NGC~4536, NGC~4569, NGC~4579, and 
NGC~4725) that were observed as part of the Herschel Reference Survey (HRS; Boselli et al.\ 2010a); and we add the SPIRE observations of M~33, from HERM33ES (Kramer et al.\ 2010). 
%
NGC~1404 and DDO~154 were observed by SPIRE but not detected, and their 
MIPS fluxes appear to be due to background sources (Dale et al.\ 2007), so we 
include these galaxies but regard their dust fluxes and masses as upper limits. 
%

The SPIRE (as well as PACS) flux densities will be shown and described in the KINGFISH 
photometry paper, Dale et al.\ (in prep.). 
For details about the other flux densities, see Dale et al.\ (2007, 2009). 
%
The galaxy distances are listed in Table~\ref{table1}, and for most of the galaxies, they are 
the same as those in Kennicutt et al.\ (in prep.). 
For a description of the distance indicators, and references for the distance measurements, 
see Kennicutt et al.\ (2003; in prep.). 
%
Some of the KINGFISH galaxies have been examined in detail in recent papers, such as NGC~1097 
(Beir\~{a}o et al.\ 2010; Sandstrom et al.\ 2010), NGC~3077 (Walter et al.\ 2011), NGC~6946 (Murphy et al.\ 2011b), and NGC~1291 (Hinz et al., in prep.). 

Elliptical apertures are used for the photometry and are chosen to approximately 
encompass all of the optical and infrared emission of a galaxy. 
Typically, this means that the $3.6~\mu$m image was used to create the aperture, since 
$3.6~\mu$m is the bandpass within which \textit{Spitzer} is most sensitive and the stellar 
disk is most spatially extended, although in a few cases the far-IR $160~\mu$m disk is more extended. 
The same aperture was used at all wavelengths. 
The global flux densities exclude foreground stars and background galaxies. 
We estimate that uncertainties involving the apertures may introduce up to 
0.1~dex errors to the flux densities; however, these errors are quite small 
and do not significantly affect the dust/stellar flux ratios used for our analysis 
(see Section~\ref{dusttostellar}).  
For details, we refer the reader to Dale et al.\ (2007 and 2009, Tables 1). 

Finally, for most of the KINGFISH galaxies that are also in SINGS, we measured 
H$\alpha$ fluxes. 
%
For these, we used H$\alpha$ images that were obtained as part of the SINGS ancillary 
program, either at the 2.1~m Kitt Peak National Observatory (KPNO) telescope or at 
the 1.5~m Cerro Tololo Inter-American Observatory (CTIO) telescope (Kennicutt et al.\ 2003). 
As described by Calzetti et al.\ (2007), exposure times were typically around 1800~s 
and standard reduction procedures were applied. 
%
There were 13 additional galaxies for which we 
did not have H$\alpha$ fluxes but which were included in Kennicutt et al.\ (2008), and we 
used their fluxes for these galaxies, which are marked in Table~\ref{table1} (with the superscript $^e$). 
For the galaxies whose H$\alpha$ fluxes we could compare, 
our fluxes are slightly higher than those of Kennicutt et al., by $\approx0.17$ dex, on average. 
For NGC~598, we obtained H$\alpha$ data from Massey et al.\ (2006), 
from which we measured a flux density of $2.4\pm0.2\times10^{-10}\,\mathrm{erg}\,\mathrm{cm}^{-2}\,\mathrm{s}^{-1}$, which is slightly lower (by $0.15$ dex) than the value given by Kennicutt et al. 
For NGC~2146, which was in neither sample, we use the H$\alpha$ flux measured by Marcum et al.\ (2001).


\subsection{Morphological Classifications}\label{morph}

The morphological types and metallicities of galaxies play important 
roles in galaxy evolution (e.g., Calura et al.\ 2008; Fontanot et al.\ 2009), 
and they have been found to be correlated with the dust and stellar 
properties of galaxies (e.g., Draine et al.\ 2007).
Nonetheless, morphologies and metallicities are notoriously difficult to 
accurately determine without significant biases, and at fixed stellar mass, 
galaxies still have a fairly wide distribution of morphologies (e.g., 
Bamford et al.\ 2009) and metallicities (e.g., Tremonti et al.\ 2004). 
In light of this, in Section~\ref{results}, we will analyze the dust/stellar flux ratios 
of the KINGFISH galaxies as a function of other 
galaxy properties, but we will mark the galaxies by their morphological 
types (described in this section) and metallicities (described in Section~\ref{metal}) such that trends for late-types and early-types, 
and metal-poor and metal-rich galaxies, can be distinguished.  Even if 
particular galaxies are misclassified or have ill-determined metallicities, 
we expect the relative morphologies and metallicities within the sample to 
be sufficiently accurate for statistical purposes.

All but four of the KINGFISH galaxies were also in the SINGS sample 
(Kennicutt et al.\ 2003), in which galaxies were selected to span a wide 
range of RC3 (de Vaucouleurs et al.\ 1991) morphological types.  Nonetheless, 
many more data have been accumulated about the nearby galaxies since these 
classifications were made, and the classifications can now be done more 
accurately and homogeneously.

Buta et al.\ (2010) have recently classified a subset of the objects in the 
\textit{Spitzer} Survey of Stellar Structure in Galaxies (Sheth et al.\ 2010), 
using 3.6$\mu$m images with good spatial resolution.  29 of these objects 
are in the KINGFISH sample, and for these galaxies, we use the updated 
morphologies.  In Table~\ref{table1}, we note the 17 galaxies which have been 
classified slightly differently than previously.  Some galaxies 
have been found to have slightly earlier types than previously; NGC~584 and 
NGC~855---ellipticals which now have faintly detected disks---are exceptions, 
as is NGC~1482, which now appears to have more of the structure of an Sa, 
rather than an S0. 
In any case, the previous RC3 classifications and Buta et al.\ (2010) classifications 
are generally consistent.  
In addition, examining many of the same galaxies, Kendall et al.\ (2011) recently 
found that spiral structures are usually similar in optical and IR images. 
We divide the KINGFISH sample into three types: there are 17 dwarf and irregular 
galaxies (Sd and later type), 32 spirals (Sa to Scd), and 10 early-types (E and S0).

Galactic bars may also be an important property, being related to gas 
concentration, star formation, and dust heating in the central regions of 
galaxies (Sheth et al.\ 2005; Engelbracht et al.\ 2010).
Approximately half of the KINGFISH galaxies have strong bars (SAB and SB), 
although because of the crudeness of the bar strength classifications, it 
is difficult to robustly determine the bar dependence of the galaxy 
properties. 
When significant, we quote the bar dependence of the galaxies' dust and stellar properties 
in the Section~\ref{results}; the dependence is usually modest at most. 

\subsection{Gas-Phase Metallicities}\label{metal}

In order to quantify the 
metallicity dependence of correlations with the dust/stellar flux ratios, we use 
the oxygen abundances measured by Moustakas et al.\ (2010).  In particular, 
we have chosen to use the abundances based on the theoretical strong-line 
calibration of Kobulnicky \& Kewley (2004), even though it yields 
overestimates for some galaxies.  The empirical calibration of Pilyugin \& 
Thuan (2005) is more accurate for typical $L_\ast$ galaxies, though it 
yields underestimates for star-forming metal-rich galaxies; more 
importantly, it was calibrated using only HII regions in spiral and 
irregular galaxies, and it may be dangerous to extrapolate beyond this regime. 
(For a discussion of the effects of using different metallicity calibrations, see e.g., Kewley \& Ellison 2008; Calura et al.\ 2009.) 
We use the `characteristic' (globally-averaged) metallicities of the galaxies, even for 
those with metallicity gradients, such as NGC~5457. 

Engelbracht et al.\ (2005) found that there appears to be an oxygen abundance 
threshold at $12+\mathrm{log}\,(O/H)\sim8.2$, such that galaxies below this 
threshold tend to have weak PAH emission; 
Draine et al.\ (2007) similarly found low PAH mass fractions in metal-poor galaxies. 
In order to determine how such 
metallicity transition may be related to the dust and stellar properties 
of galaxies, we use only the relative abundances of Moustakas et al.\ (2010), 
and split the KINGFISH sample into thirds, consisting of 
`metal-rich' (i.e., highest $O/H$ metallicity), `intermediate' metallicity, and 
`metal-poor' (i.e., lowest metallicity) galaxies.  Our results are not 
significantly dependent on which metallicity calibration we use; the Pilyugin \& Thuan 
(2005) calibration yields similar \textit{relative} abundances for most of the 
sample.  For the galaxies lacking prominent emission lines, we estimate 
approximate metallicities based on the $B$-band $L-Z$ relation (Moustakas et 
al. 2010; \textit{cf}., Tremonti et al.\ 2004).  Note that this relation 
has substantial scatter for faint galaxies, and this is another motivation 
for focusing on the relative abundances. 

The absolute metallicity 
scales we use to split the sample are $12+\mathrm{log}\,(O/H)=8.88$ and 
9.08 (or 8.29 and 8.42 using the Pilyugin \& Thuan (2005) calibration), 
but these absolute abundances should be treated with caution.  When we refer 
to `metal-poor' and `metal-rich' galaxies, these are meant to be relative to 
other KINGFISH galaxies.


\section{Dust/Stellar Flux Ratio}\label{dusttostellar}

\subsection{Motivation}

Our goal is to estimate the emission from stars and dust 
in a way that is empirical and as model-independent as possible.  This 
allows us to exploit the diversity of data that have been accumulated for these 
galaxies, and to compare to methods involving SED models (e.g., Draine et al.\ 
2007; da Cunha et al.\ 2010). 
As described in detail below, we compute a dust/stellar flux ratio for each galaxy 
by integrating the SED from mid-IR to submm wavelengths and from UV to mid-IR wavelengths, and then taking the ratio.  
This is a quantity that can be 
physically interpreted as the amount of emission being reprocessed by 
dust grains (mostly small and large grains), 
relative to the unobscured emission from stars (especially young massive O and B stars, as well as intermediate-age AGB stars).

Galaxy SEDs, and dust/stellar flux ratios in particular, are related to other properties 
indicative of a galaxy's evolution, such as metallicity and morphology 
(e.g., Groves et al.\ 2008; Fontanot et al.\ 2009), which were discussed in Sections~\ref{morph} and Section~\ref{metal}. 
Note that the dust/stellar flux is similar, but not equivalent, to the dust/stellar \textit{mass} ratio, discussed in Section~\ref{masses}. 
The dust/stellar flux ratio is also similar to the dust/FUV or IR/FUV ratio studied by many authors 
(e.g., Meurer et al.\ 1999; Kong et al.\ 2004; Johnson et al.\ 2007; Boquien et al.\ 2009; Wijesinghe et al.\ 2011), but these quantities specifically measure the attenuation of UV photons, while the dust/stellar flux ratio also accounts for dust absorbing optical photons from older stellar populations. 
A galaxy's dust/stellar flux is related to its specific star formation rate, 
as we will show in Section~\ref{SFRsec}, and its star formation history. 

Lastly, note that galaxy geometry (i.e., inclination) and differential extinction will affect these ratios within galaxies to some degree (e.g., Jonsson et al.\ 2010). 
A few galaxies in the sample are highly inclined, such as NGC~4594 and NGC~4631, but as we will show in the next section, their dust/stellar flux ratios do not appear to be biased (with suppressed stellar emission).  Similarly, Dale et al.\ (2007) examined the IR/UV luminosity ratio of SINGS galaxies and did not detect a significant trend with disk inclination.

\subsection{Procedure}

Our procedure is as follows.  We begin by compiling the flux densities from the 
UV to far-infrared from Dale et al.\ (2007, 2009).  For the $ugriz$-band optical 
SDSS data obtained for 11 of the 17 LVL galaxies in KINGFISH, and the $BVRI$-band 
data obtained for the other galaxies, we first convert the apparent magnitudes to 
flux densities.\footnote{For the conversion from mag to Jy, 
we used the following conversion for the $BVRI$-band magnitudes: \texttt{http://ssc.spitzer.caltech.edu/tools/magtojy/ref.html}; and the following for the $ugriz$-band magnitudes: \texttt{http://www.sdss.org/dr5/algorithms/fluxcal.html}.}  
We then use the flux densities in units of Janskys, 
at wavelengths ranging from 0.15 to 850 $\mu$m. 
We have measured the SPIRE fluxes at 250, 350, and 500 $\mu$m ourselves, 
using the same apertures that had been used at shorter wavelengths. 
The SPIRE photometry and global flux densities will be presented in Dale et al.\ (in prep.).

For galaxies missing data or detections at UV or submm wavelengths, we attempt to 
extrapolate the SEDs, in order to more consistently compare all of the galaxies in the 
sample.  The extrapolations may be uncertain, but they yield more accurate 
dust/stellar flux ratios than neglecting these regions of the SEDs.  Nonetheless, 
the effects of these extrapolations are usually extremely small, 
although as noted below, the UV extrapolations of a few galaxies have an effect of $>10\%$. 
We stress that 
\textit{the dust/stellar flux ratios are largely determined by the stellar and IR peaks of 
the SEDs, which are well determined for the 62 galaxies in our fiducial sample.} 

For galaxies with SPIRE detections but lacking detections with SCUBA at 850~$\mu$m, 
we perform a linear fit to the SPIRE log flux densities (at 250-500~$\mu$m), and 
extrapolate to 850~$\mu$m, and give this flux an uncertainty of 1.5 times the uncertainty of the 500~$\mu$m flux. 
(The choice of 1.5 times the uncertainty is arbitrary; if we were to double this uncertainty, it would not significantly affect the final uncertainty estimated for the dust/stellar flux ratio.) 
We have verified that this is an accurate extrapolation for galaxies with both SPIRE and SCUBA detections, and in 
any case, it contributes a negligible effect to the dust/stellar flux ratios: the 
ratios are affected by $0.002$~dex on average and $0.03$~dex ($6\%$) at most.

\begin{figure}
 \includegraphics[width=\hsize]{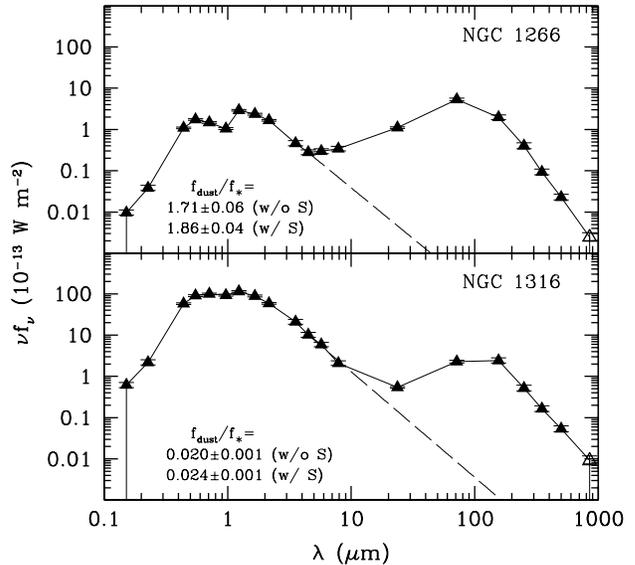} 
 \caption{Spectral energy distributions of NGC~1266 and NGC~1316. 
          The dust/stellar flux ratios of the galaxies, with and without the 
          SPIRE data included, are listed in the lower left of each panel. 
          The dashed line indicates the estimated stellar contribution 
          that is integrated at $\lambda>4.5$~$\mu$m. 
          The open points at 850~$\mu$m are extrapolated from the SPIRE flux densities. 
          See text for details.
         }
 \label{SEDs}
\end{figure}

The extrapolation at the UV end is slightly more important, because these wavelengths 
are closer to the stellar peak of the SEDs of some galaxies than 850~$\mu$m is to the 
far-IR peak. 
For the galaxies only lacking GALEX detections in the FUV, we extrapolate linearly 
from the $NUV$- and $U$-bands (or $NUV$- and $u$-bands), and give this flux 
1.5 times the uncertainty of the NUV flux density. 
Three galaxies lack UV data altogether, and we examined each of these individually. 
The turnover of NGC~3077's SED is at longer wavelengths ($\approx800~$nm), 
so we extrapolated linearly from the $u$- and $g$-bands for this galaxy and accordingly 
gave these UV fluxes larger uncertainties. 
We assume that NGC~1377's SED resembles that of similar S0's in the sample (such as NGC~1266's in Fig.~\ref{SEDs}), which turn over steeply.  
Using the mean and variance of $FUV$-$NUV$ and $NUV$-$B$ colors of these galaxies, 
we estimated the UV end of this galaxy's SED, and the uncertainty of the resulting dust/stellar flux ratio. 
Dwarf and irregular galaxies, including those in our sample, tend to have a shallower UV slope (Dale et al.\ 2007). 
Using the mean and variance of $FUV$-$NUV$ and $NUV$-$B$ of these galaxies, 
we estimated the UV end of NGC~5408's SED, and the uncertainty of the resulting dust/stellar flux ratio. 
The extrapolations only out to the bluest $FUV$ point 
have a very small effect on the dust/stellar flux 
ratios.  
The extrapolations through both UV bands are more significant; 
without them, one would overestimate the dust/stellar flux ratios of NGC~3077, NGC~1377, 
and NGC~5408 by $12\%$, $13\%$, and $50\%$, respectively. 
In any case, these factors 
are still relatively small on a log scale. 
%

In order to demarcate ``dust'' and ``stellar'' emission in the SEDs, the simplest 
approach would be to impose a strict wavelength cut at 5~$\mu$m, which is 
motivated by the fact that most stellar emission occurs at shorter wavelengths 
and dust mostly emits at longer wavelengths (e.g., Engelbracht et al.\ 2008). 
Nevertheless, a small fraction of stellar emission occurs in the mid-IR, 
and for some galaxies, and for early-types in particular, the emission from 
dust does not dominate until longer wavelengths. 
In addition, some galaxies have strong polycyclic aromatic hydrocarbon (PAH) features 
that make it difficult to determine a strict demarcation of the stellar and dust SEDs. 

Motivated by this, we attempt to estimate the stellar contribution beyond 5~$\mu$m, 
as some authors have done (e.g., Draine et al.\ 2007; Mu\~{n}oz-Mateos et al.\ 2009a). 
For most galaxies, we fit a power-law to the stellar SED from the $K$-band to 4.5~$\mu$m,  
and extrapolate to estimate this extra contribution (see Figure~\ref{SEDs}).  
Two exceptions are NGC~1377, for which we extrapolate from $K$-band to 3.6~$\mu$m, and DDO~165, for which we extrapolate from $K$ to 8~$\mu$m but exclude the 3.6 and 4.5~$\mu$m fluxes, 
as these appear to be slightly enhanced. 
This is not an ideal solution, because dust contamination may occur even at 2~$\mu$m 
(Mentuch et al.\ 2010), but for most of the galaxies in our sample, 
the Wien side of the dust SED dominates and its contribution increases with wavelength at $\lambda>4.5~\mu$m. 
The effect of accounting for the stellar contribution at these wavelengths has a very 
small effect on the dust/stellar flux ratios, and is significant only for the 
early-types (lowers their $f_\mathrm{dust}/f_\ast$ by $<0.25$ dex, and by 0.5 dex for NGC~1404) 
and IC~342, but the contribution for this galaxy is very uncertain, because of its 
uncertain near- and mid-IR fluxes. 
We assume that the non-stellar contribution at $\lambda<4.5~\mu$m is minimal 
(see Meidt et al. 2011).
%


To estimate the stellar and dust emission, which we call 
$f_\ast$ and $f_\mathrm{dust}$, 
we integrate over the SED at $\lambda\leq4.5\mu$m and add the mid-IR stellar contribution 
for the former and integrate the SED at $\lambda\geq4.5\mu$m and subtract the same 
contribution for the latter.  The area under the SED is computed directly, 
using Jy and Hz as the units. 
%
In particular, 
for each galaxy, we first linearly interpolate $\mathrm{log}\,f(\mathrm{log}\,\nu)$ 
over the range of wavelengths. 
Then we integrate over $f(\nu)$ (i.e., not in log space).
\begin{equation}
  \frac{f_\mathrm{dust}}{f_\ast} \,\equiv\,
    \frac{ \int_{\lambda=4.5\mu m}^{\lambda=850\mu m}\,{\rm d}\nu\, f_\nu \,-\, \int_{4.5\mu m}^{50\mu m}\,{\rm d}\nu\,f_\mathrm{star} } 
         { \int_{\lambda=0.15\mu m}^{\lambda=4.5\mu m}\,{\rm d}\nu\, f_\nu \,+\, \int_{4.5\mu m}^{50\mu m}\,{\rm d}\nu\,f_\mathrm{star} } \, , 
 \label{ratio}
\end{equation}
where $f_\mathrm{star}$ is the estimated stellar contribution at $\lambda\geq4.5\mu$m, 
and is usually relatively small. 
Note that $f_\mathrm{dust}$ is closely related to what some call total infrared luminosity, 
which we discuss in Section~\ref{TIRlum}. 
%
To estimate the uncertainties, 
we simply assume that the errors of the flux densities have a Gaussian 
distribution, sample from these distributions 10,000 times, and compute the 
variance around the mean dust/stellar flux ratio.
Because the stellar and far-IR SED peaks are usually well determined for these galaxies, 
the uncertainties of $f_\mathrm{dust}/f_\ast$ are usually small, 
even for the few galaxies with gaps in their SEDs, such as NGC~5408.

\subsection{Resulting Dust/Stellar Flux Ratios}

Two example SEDs are shown in Figure~\ref{SEDs}.  NGC~1266 and NGC~1316 are both S0s, 
and as we shall show later, they have similar masses.  Nevertheless, they have very 
different dust/stellar flux ratios.  In addition, they are among the galaxies that, 
before \textit{Herschel}, did not have any far-IR data beyond $160~\mu$m. 

\begin{figure}
 \includegraphics[width=\hsize]{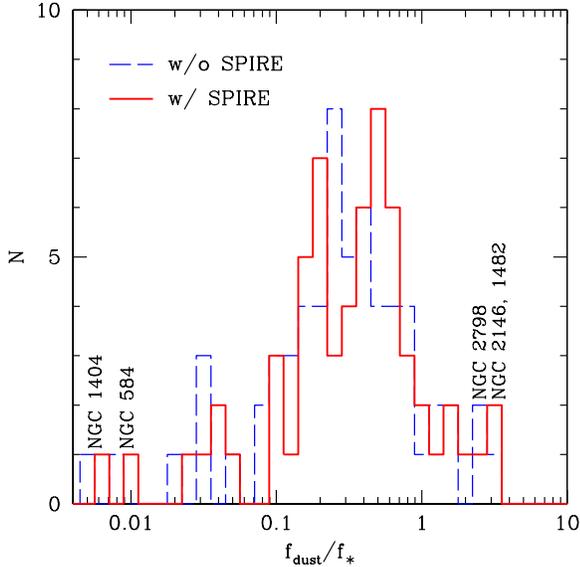} 
 \caption{Distribution of dust/stellar emission ratios of the KINGFISH galaxies.  
          The histograms show the distributions without (blue dashed line) and with 
          (red solid line) the SPIRE fluxes included.}
 \label{fracdist}
\end{figure}

The distribution of dust/stellar emission for our fiducial sample of 62 
galaxies is shown in Figure~\ref{fracdist}.
For comparison, we also show the $f_\mathrm{dust}/f_\ast$ distribution without the 
SPIRE data (and without the $850~\mu$m extrapolation). 
In other words, we compare to the 
distribution one would have obtained before \textit{Herschel}, where the only far-IR data 
were from MIPS (and IRAS and SCUBA, for a few galaxies). 

For some galaxies, especially those with SEDs peaking at long wavelengths, 
\textit{Herschel} contributes stronger constraints on the ratio 
of dust/stellar emission, by tracing additional cold dust components 
not detected by \textit{Spitzer}. 
Consequently, by adding a contribution in the far-IR, the SPIRE fluxes slightly 
increase the dust/stellar flux ratio of some galaxies in the sample.  
The mean ratio is 
$\overline{\mathrm{log}\, f_\mathrm{dust}/f_\ast}=-0.59\pm0.07$ without SPIRE and 
$\overline{\mathrm{log}\, f_\mathrm{dust}/f_\ast}=-0.52\pm0.07$ with SPIRE, 
an increase of about 17$\%$.\footnote{The mean ratios, without using logarithms, are $\overline{f_\mathrm{dust}/f_\ast}=0.51\pm0.08$ without SPIRE and $\overline{f_\mathrm{dust}/f_\ast}=0.55\pm0.09$ with SPIRE, an increase of $\approx8\%$. Since $f_\mathrm{dust}/f_\ast$ spans three orders of magnitude, however, we use the logs of the ratios throughout this paper.} 
(Note that the standard deviations about these means are $\approx0.54$~dex.) 
NGC~1512 is the galaxy for which the inclusion of SPIRE data has the largest effect, 
increasing its dust/stellar flux ratio by 0.23~dex.
The mean ratio varies with morphology as well, 
as shown in Table~\ref{table2} and discussed in Section~\ref{results}. 

The three galaxies with the largest dust/stellar flux ratios are NGC~1482, NGC~2146, and NGC~2798, 
which are all starbursting early-type spirals.  
NGC~1482 also has evidence of superwind outflows (Hota \& Saikia 2005), 
while NGC~2146 and NGC~2798 have peculiar morphologies, possibly indicative of interactions or mergers (but see Greve et al.\ 2006 on NGC~2146). 
The two galaxies with the smallest ratios are NGC~1404 and NGC~584, 
which are both massive ellipticals (or E/S0) with very little FIR emission; NGC~1404 is also 
experiencing ram-pressure stripping as it falls through the Fornax cluster (Machacek et al.\ 2005). 
%
NGC~1404's $f_\mathrm{dust}/f_\ast$ ($\approx0.006$) should be considered an upper limit, because 
its FIR and submm fluxes appear to be due to a background source (Dale et al., in prep.), 
and it is an indication of the most stellar-dominated a galaxy's SED can be.

In Table~\ref{table1}, we show the dust/stellar flux ratios of the KINGFISH galaxies.  
The near-IR morphologies, TIR luminosities, dust masses, stellar masses, and star 
formation rates  are also listed in the table, and are discussed below.


\begin{table*}
 \caption{KINGFISH Galaxy Properties}
 \centering
 \begin{tabular}{ l | l l c c c c c c c }
   \hline
   Galaxy & Type & Dist. & $f_\mathrm{dust}/f_\ast$ & log~$L_\mathrm{TIR}$ & $T_\mathrm{dust}$ & log~$M_\mathrm{dust}$ & log~$M_\ast$ & log~SFR$_{\mathrm{FUV}+\mathrm{TIR}}$ & log~SFR$_{H\alpha+24\mu {\rm m}}$ \\
    & & (Mpc) & & (log~erg~$s^{-1}$) & (K) & (log~$M_\odot$) & (log~$M_\odot$) & (log~$M_\odot~\mathrm{yr}^{-1}$) & (log~$M_\odot~\mathrm{yr}^{-1}$) \\
   \hline
   NGC 0337 & SABcdp$^{\rm a}$            & 22.9  & 0.77 $\pm$ 0.01   & 43.84 $\pm$ 0.04 & 28.1 $\pm$ 0.7 & 7.07 $\pm$ 0.08           &  9.47 $\pm$ 0.22           &  0.29 $\pm$ 0.04           &  0.24 $\pm$ 0.05 \\
   NGC 0584 & S\underline{A}B0$-^{\rm a}$ & 20.8  & 0.009 $\pm$ 0.001 & 42.60 $\pm$ 0.06 & 24.5 $\pm$ 0.6 & 5.58 $\pm$ 0.15           & 11.12 $\pm$ 0.07           & -1.01 $\pm$ 0.05           &     \nodata      \\
   NGC 0628 & SAc                         &  7.3  & 0.50 $\pm$ 0.01   & 43.50 $\pm$ 0.05 & 24.0 $\pm$ 0.6 & 7.03 $\pm$ 0.08           &  9.57 $\pm$ 0.13           &  0.04 $\pm$ 0.05           & -0.01 $\pm$ 0.05 \\
   NGC 0855 & SA0$-^{\rm a}$              &  9.73 & 0.26 $\pm$ 0.001  & 42.21 $\pm$ 0.04 & 28.5 $\pm$ 0.9 & 5.49 $\pm$ 0.08           &  8.67 $\pm$ 0.10           & -1.32 $\pm$ 0.04           & -1.34 $\pm$ 0.05$^{\rm e}$ \\
   NGC 0925 & SABd                        &  9.04 & 0.36 $\pm$ 0.01   & 43.25 $\pm$ 0.04 & 23.7 $\pm$ 0.5 & 6.98 $\pm$ 0.08           &  9.48 $\pm$ 0.14           & -0.11 $\pm$ 0.05           & -0.22 $\pm$ 0.05 \\
   NGC 1097 & SBa\underline{b}p           & 19.09 & 0.65 $\pm$ 0.01   & 44.52 $\pm$ 0.04 & 26.2 $\pm$ 0.6 & 7.80 $\pm$ 0.08           & 10.74 $\pm$ 0.12           &  0.91 $\pm$ 0.03           &  0.94 $\pm$ 0.05 \\
   NGC 1266 & SB0                         & 30.6  & 1.86 $\pm$ 0.04   & 44.00 $\pm$ 0.03 & 36.0 $\pm$ 1.0 & 6.66 $\pm$ 0.08           & 10.14 $\pm$ 0.12           &  0.31 $\pm$ 0.03           &  0.36 $\pm$ 0.05 \\
   NGC 1291 & SAB0$+$                     & 10.4  & 0.043 $\pm$ 0.001 & 43.12 $\pm$ 0.06 & 22.4 $\pm$ 0.5 & 6.76 $\pm$ 0.08           & 10.79 $\pm$ 0.10           & -0.45 $\pm$ 0.04           & -0.07 $\pm$ 0.05 \\
   NGC 1316 & SAB0                        & 20.1  & 0.024 $\pm$ 0.001 & 43.55 $\pm$ 0.04 & 26.8 $\pm$ 0.7 & 6.79 $\pm$ 0.08           & 11.42 $\pm$ 0.09           & -0.07 $\pm$ 0.04           & -0.42 $\pm$ 0.07 \\
   NGC 1377 & S0                          & 24.6  & 1.69 $\pm$ 0.02   & 43.74 $\pm$ 0.02 & 43.5 $\pm$ 1.8 & 5.78 $\pm$ 0.09           &  9.28 $\pm$ 0.14           &  0.05 $\pm$ 0.04$^{\rm d}$ &  0.45 $\pm$ 0.06 \\
   NGC 1404 & E                           & 19.5  &     $<$0.006      & 42.58 $\pm$ 0.05 &    \nodata     & 6.5  $\pm$ 1.6$^{\rm b}$  & 10.85 $\pm$ 0.13           & -0.94 $\pm$ 0.05           & -0.60 $\pm$ 0.05 \\
   NGC 1482 & Sa$^{\rm a}$                & 22.6  & 3.37 $\pm$ 0.07   & 44.29 $\pm$ 0.03 & 31.8 $\pm$ 0.9 & 7.13 $\pm$ 0.08           &  9.99 $\pm$ 0.11           &  0.60 $\pm$ 0.03           &  0.71 $\pm$ 0.08 \\
   NGC 1512 & SBa$^{\rm a}$               & 14.35 & 0.28 $\pm$ 0.01   & 43.35 $\pm$ 0.05 & 20.9 $\pm$ 0.8 & 7.00 $\pm$ 0.08           & 10.10 $\pm$ 0.11           & -0.09 $\pm$ 0.05           &  0.23 $\pm$ 0.05 \\
   Ho II    & Im                          &  3.6  & 0.09 $\pm$ 0.001  & 41.63 $\pm$ 0.04 & 36.5 $\pm$ 1.1 & 4.05 $\pm$ 0.20           &  7.73 $\pm$ 0.15           & -1.14 $\pm$ 0.06           & -1.17 $\pm$ 0.06 \\
   DDO 053  & Im                          &  3.6  & 0.27 $\pm$ 0.01   & 40.71 $\pm$ 0.05 & 30.5 $\pm$ 0.9 & 4.01 $\pm$ 0.10           &  6.35 $\pm$ 0.20           & -2.34 $\pm$ 0.06           &     \nodata      \\
   NGC 2798 & S\underline{A}Bap           & 25.8  & 2.54 $\pm$ 0.04   & 44.18 $\pm$ 0.03 & 34.9 $\pm$ 1.1 & 6.83 $\pm$ 0.08           & 10.04 $\pm$ 0.13           &  0.50 $\pm$ 0.03           &  0.75 $\pm$ 0.05 \\
   NGC 2841 & S\underline{A}Ba$^{\rm a}$  & 14.1  & 0.15 $\pm$ 0.002  & 43.72 $\pm$ 0.05 & 22.1 $\pm$ 0.4 & 7.34 $\pm$ 0.08           & 10.17 $\pm$ 0.14           &  0.12 $\pm$ 0.04           &     \nodata      \\
   NGC 2915 & I0                          &  3.78 & 0.10 $\pm$ 0.001  & 41.27 $\pm$ 0.04 & 28.9 $\pm$ 0.9 & 4.59 $\pm$ 0.08           &  7.57 $\pm$ 0.20           & -1.57 $\pm$ 0.06           & -1.75 $\pm$ 0.06$^{\rm e}$ \\
   Ho I     & IABm                        &  3.6  & 0.11 $\pm$ 0.001  & 40.79 $\pm$ 0.07 & 26.2 $\pm$ 0.9 & 4.54 $\pm$ 0.08           &  6.80 $\pm$ 0.22           & -2.08 $\pm$ 0.06           &     \nodata     \\
   NGC 2976 & SABd$^{\rm a}$              &  3.6  & 0.47 $\pm$ 0.01   & 42.54 $\pm$ 0.05 & 25.9 $\pm$ 0.7 & 5.97 $\pm$ 0.08           &  8.97 $\pm$ 0.11$^{\rm c}$ & -1.00 $\pm$ 0.04           & -0.97 $\pm$ 0.05 \\
   NGC 3049 & SBab                        & 19.2  & 0.63 $\pm$ 0.01   & 43.16 $\pm$ 0.03 & 27.5 $\pm$ 0.7 & 6.45 $\pm$ 0.08           &  8.58 $\pm$ 0.04           & -0.47 $\pm$ 0.04$^{\rm d}$ & -0.27 $\pm$ 0.06 \\
   NGC 3077 & I0p                         &  3.6  & 0.30 $\pm$ 0.001  & 42.46 $\pm$ 0.04 & 30.1 $\pm$ 0.9 & 5.52 $\pm$ 0.08           &  9.29 $\pm$ 0.07$^{\rm c}$ & -1.21 $\pm$ 0.05$^{\rm d}$ & -1.23 $\pm$ 0.05$^{\rm e}$ \\
   M81 dwB  & Im                          &  3.6  & 0.17 $\pm$ 0.01   & 40.40 $\pm$ 0.14 & 25.0 $\pm$ 0.7 & 4.06 $\pm$ 0.09           &  6.36 $\pm$ 0.20           & -2.82 $\pm$ 0.06           &     \nodata      \\
   NGC 3190 & SAap                        & 19.3  & 0.19 $\pm$ 0.002  & 43.49 $\pm$ 0.04 & 25.2 $\pm$ 0.5 & 6.89 $\pm$ 0.08           & 10.03 $\pm$ 0.14           & -0.20 $\pm$ 0.04           & -0.45 $\pm$ 0.05 \\
   NGC 3184 & SAbc$^{\rm a}$              &  8.7  & 0.32 $\pm$ 0.01   & 43.40 $\pm$ 0.05 & 23.4 $\pm$ 0.5 & 6.90 $\pm$ 0.08           &  9.24 $\pm$ 0.17           & -0.08 $\pm$ 0.04           & -0.25 $\pm$ 0.05 \\
   NGC 3198 & SABbc$^{\rm a}$             & 14.5  & 0.43 $\pm$ 0.01   & 43.60 $\pm$ 0.05 & 23.6 $\pm$ 0.5 & 7.18 $\pm$ 0.08           &  9.85 $\pm$ 0.11           &  0.12 $\pm$ 0.04           & -0.03 $\pm$ 0.05 \\
   IC 2574  & IBm$^{\rm a}$               &  3.6  & 0.18 $\pm$ 0.001  & 41.91 $\pm$ 0.04 & 25.9 $\pm$ 0.6 & 5.57 $\pm$ 0.08           &  8.16 $\pm$ 0.20           & -1.11 $\pm$ 0.06           & -1.72 $\pm$ 0.05 \\
   NGC 3265 & E                           & 19.6  & 0.97 $\pm$ 0.02   & 43.05 $\pm$ 0.03 & 31.8 $\pm$ 0.9 & 6.00 $\pm$ 0.08           &  8.70 $\pm$ 0.12           & -0.60 $\pm$ 0.03           & -0.53 $\pm$ 0.05 \\
   NGC 3351 & SBa$^{\rm a}$               &  9.8  & 0.36 $\pm$ 0.01   & 43.51 $\pm$ 0.05 & 25.6 $\pm$ 0.6 & 6.87 $\pm$ 0.08           & 10.28 $\pm$ 0.12$^{\rm c}$ & -0.07 $\pm$ 0.04           & -0.49 $\pm$ 0.09$^{\rm e}$ \\
   NGC 3521 & SABbc                       & 12.44 & 0.53 $\pm$ 0.02   & 44.24 $\pm$ 0.05 & 24.9 $\pm$ 0.6 & 7.63 $\pm$ 0.08           & 10.78 $\pm$ 0.12$^{\rm c}$ &  0.59 $\pm$ 0.04           &  0.59 $\pm$ 0.05 \\
   NGC 3621 & SAd                         &  6.9  & 0.54 $\pm$ 0.01   & 43.57 $\pm$ 0.04 & 25.4 $\pm$ 0.6 & 6.97 $\pm$ 0.08           &  9.43 $\pm$ 0.11           &  0.06 $\pm$ 0.04           & -0.03 $\pm$ 0.05$^{\rm e}$ \\
   NGC 3627 & SBbp                        & 10.3  & 0.60 $\pm$ 0.02   & 44.15 $\pm$ 0.04 & 27.2 $\pm$ 0.7 & 7.32 $\pm$ 0.08           & 10.57 $\pm$ 0.13$^{\rm c}$ &  0.50 $\pm$ 0.04           &  0.49 $\pm$ 0.05 \\
   NGC 3773 & SA0                         & 12.4  & 0.46 $\pm$ 0.01   & 42.44 $\pm$ 0.04 & 30.2 $\pm$ 0.8 & 5.44 $\pm$ 0.08           &  8.31 $\pm$ 0.16           & -0.91 $\pm$ 0.05           & -0.89 $\pm$ 0.05 \\
   NGC 3938 & SAc                         & 12.1  & 0.45 $\pm$ 0.01   & 43.57 $\pm$ 0.04 & 24.8 $\pm$ 0.5 & 6.94 $\pm$ 0.08           &  9.12 $\pm$ 0.04           & -0.05 $\pm$ 0.05$^{\rm d}$ & -0.05 $\pm$ 0.05 \\
   NGC 4236 & SBdm                        &  3.6  & 0.14 $\pm$ 0.001  & 42.09 $\pm$ 0.05 & 25.0 $\pm$ 0.7 & 5.83 $\pm$ 0.08           &  8.18 $\pm$ 0.13           & -0.93 $\pm$ 0.06           & -0.97 $\pm$ 0.05$^{\rm e}$ \\
   NGC 4254 & SAcp                        & 15.3  & 0.81 $\pm$ 0.01   & 44.29 $\pm$ 0.04 & 25.5 $\pm$ 0.5 & 7.56 $\pm$ 0.08           &  9.61 $\pm$ 0.15           &  0.63 $\pm$ 0.05$^{\rm d}$ &  0.72 $\pm$ 0.05 \\
   NGC 4321 & SABbc                       & 15.3  & 0.59 $\pm$ 0.01   & 44.22 $\pm$ 0.04 & 24.4 $\pm$ 0.5 & 7.61 $\pm$ 0.08           & 10.36 $\pm$ 0.11           &  0.58 $\pm$ 0.05$^{\rm d}$ &  0.45 $\pm$ 0.05 \\
   NGC 4536 & SAB\underline{b}c           & 15.3  & 0.77 $\pm$ 0.02   & 44.00 $\pm$ 0.03 & 26.9 $\pm$ 0.6 & 7.28 $\pm$ 0.08           &  9.49 $\pm$ 0.11           &  0.39 $\pm$ 0.03           &  0.39 $\pm$ 0.05 \\
   NGC 4559 & SBcd                        &  8.45 & 0.34 $\pm$ 0.01   & 43.28 $\pm$ 0.04 & 24.5 $\pm$ 0.5 & 6.83 $\pm$ 0.08           &  8.93 $\pm$ 0.20           & -0.11 $\pm$ 0.04           &     \nodata      \\
   NGC 4569 & SABab                       & 15.3  & 0.20 $\pm$ 0.002  & 43.72 $\pm$ 0.04 & 24.0 $\pm$ 0.5 & 7.16 $\pm$ 0.08           & 10.38 $\pm$ 0.12           &  0.09 $\pm$ 0.04           &     \nodata      \\
   NGC 4579 & SBa$^{\rm a}$               & 15.3  & 0.15 $\pm$ 0.01   & 43.64 $\pm$ 0.05 & 23.4 $\pm$ 0.5 & 7.12 $\pm$ 0.08           &  9.96 $\pm$ 0.23           &  0.01 $\pm$ 0.04           & -0.16 $\pm$ 0.05 \\
   NGC 4594 & SAa                         &  9.4  & 0.035 $\pm$ 0.001 & 43.26 $\pm$ 0.05 & 22.1 $\pm$ 0.4 & 6.91 $\pm$ 0.08           & 11.06 $\pm$ 0.12           & -0.38 $\pm$ 0.04           & -0.55 $\pm$ 0.05 \\
   NGC 4625 & SABm                        &  9.3  & 0.37 $\pm$ 0.01   & 42.40 $\pm$ 0.05 & 24.8 $\pm$ 0.6 & 5.89 $\pm$ 0.08           &  8.72 $\pm$ 0.14$^{\rm c}$ & -0.97 $\pm$ 0.05           & -1.31 $\pm$ 0.07$^{\rm e}$ \\
   NGC 4631 & SBd                         &  7.62 & 1.11 $\pm$ 0.01   & 44.00 $\pm$ 0.05 & 27.7 $\pm$ 0.8 & 7.26 $\pm$ 0.08           &  9.76 $\pm$ 0.14$^{\rm c}$ &  0.43 $\pm$ 0.04           &  0.37 $\pm$ 0.05 \\
   NGC 4725 & SABa$^{\rm a}$              & 12.7  & 0.16 $\pm$ 0.01   & 43.61 $\pm$ 0.04 & 21.1 $\pm$ 0.4 & 7.34 $\pm$ 0.08           & 10.58 $\pm$ 0.12           &  0.07 $\pm$ 0.03           & -0.09 $\pm$ 0.04 \\
   NGC 4736 & S\underline{A}Ba$^{\rm a}$  &  4.66 & 0.25 $\pm$ 0.002  & 43.41 $\pm$ 0.04 & 29.3 $\pm$ 0.8 & 6.52 $\pm$ 0.08           & 10.34 $\pm$ 0.13$^{\rm c}$ & -0.16 $\pm$ 0.04           & -0.66 $\pm$ 0.07$^{\rm e}$ \\
   DDO 154  & IBm                         &  4.3  &     $<$0.055      & 40.43 $\pm$ 0.13 &    \nodata     & 4.8 $\pm$ 0.6$^{\rm b}$   &  6.63 $\pm$ 0.20           & -2.04 $\pm$ 0.06           & -2.74 $\pm$ 0.14 \\
   NGC 4826 & SAab                        &  5.57 & 0.18 $\pm$ 0.001  & 43.31 $\pm$ 0.04 & 29.1 $\pm$ 0.8 & 6.38 $\pm$ 0.08           &  9.99 $\pm$ 0.12           & -0.34 $\pm$ 0.04           & -0.73 $\pm$ 0.05 \\
   DDO 165  & Im                          &  3.6  & 0.043 $\pm$ 0.001 & 40.40 $\pm$ 0.20 & 23.5 $\pm$ 1.1 & 4.19 $\pm$ 0.10           &  6.83 $\pm$ 0.40           & -2.03 $\pm$ 0.07           & -2.61 $\pm$ 0.07 \\
   NGC 5055 & SAbc$^{\rm a}$              & 10.16 & 0.47 $\pm$ 0.01   & 44.14 $\pm$ 0.06 & 24.1 $\pm$ 0.5 & 7.61 $\pm$ 0.08           & 10.76 $\pm$ 0.12$^{\rm c}$ &  0.52 $\pm$ 0.04           &  0.34 $\pm$ 0.08$^{\rm e}$ \\
   NGC 5398 & SBdm                        &  8.33 & 0.30 $\pm$ 0.002  & 42.25 $\pm$ 0.04 & 27.3 $\pm$ 0.7 & 5.59 $\pm$ 0.08           &  7.86 $\pm$ 0.10           & -1.05 $\pm$ 0.05           & -1.07 $\pm$ 0.05 \\
   NGC 5408 & IBm                         &  4.8  & 0.20 $\pm$ 0.02  & 41.88 $\pm$ 0.03 & 25.7 $\pm$ 1.1 & 4.68 $\pm$ 0.08           &  8.29 $\pm$ 0.15           & -1.29 $\pm$ 0.19$^{\rm d}$ & -1.04 $\pm$ 0.03$^{\rm e}$ \\
   NGC 5457 & Sc                          &  7.1  & 0.49 $\pm$ 0.01   & 44.01 $\pm$ 0.05 & 24.3 $\pm$ 0.6 & 7.52 $\pm$ 0.08           & 10.03 $\pm$ 0.06$^{\rm c}$ &  0.60 $\pm$ 0.05           &  0.27 $\pm$ 0.14$^{\rm e}$ \\
   NGC 5474 & SAcd                        &  6.8  & 0.20 $\pm$ 0.002  & 42.33 $\pm$ 0.06 & 24.6 $\pm$ 0.6 & 6.00 $\pm$ 0.08           &  8.70 $\pm$ 0.11$^{\rm c}$ & -0.79 $\pm$ 0.06           & -0.96 $\pm$ 0.07 \\
   NGC 5713 & SBabp$^{\rm a}$             & 21.37 & 1.31 $\pm$ 0.01   & 44.14 $\pm$ 0.03 & 30.0 $\pm$ 0.8 & 7.07 $\pm$ 0.08           & 10.07 $\pm$ 0.11           &  0.48 $\pm$ 0.03           &  0.46 $\pm$ 0.05 \\
   NGC 5866 & S0                          & 15.3  & 0.11 $\pm$ 0.001  & 43.38 $\pm$ 0.04 & 27.9 $\pm$ 0.7 & 6.57 $\pm$ 0.08           & 10.02 $\pm$ 0.09           & -0.30 $\pm$ 0.04           &     \nodata      \\
   NGC 6946 & SABcd                       &  6.8  & 0.60 $\pm$ 0.001  & 44.17 $\pm$ 0.04 & 26.0 $\pm$ 0.6 & 7.47 $\pm$ 0.08           &  9.96 $\pm$ 0.40           &  0.60 $\pm$ 0.04           &  0.11 $\pm$ 0.04$^{\rm e}$ \\
   NGC 7331 & SAb                         & 14.9  & 0.59 $\pm$ 0.001  & 44.37 $\pm$ 0.04 & 26.1 $\pm$ 0.6 & 7.71 $\pm$ 0.08           & 10.58 $\pm$ 0.12           &  0.71 $\pm$ 0.04           &  0.60 $\pm$ 0.05 \\
   NGC 7793 & SAc$^{\rm a}$               &  3.91 & 0.39 $\pm$ 0.002  & 42.91 $\pm$ 0.06 & 24.1 $\pm$ 0.6 & 6.51 $\pm$ 0.08           &  9.00 $\pm$ 0.16           & -0.44 $\pm$ 0.05           & -0.62 $\pm$ 0.05 \\
   IC 342   & SABcd                       &  3.28 & 1.5  $\pm$ 0.2    & 43.95 $\pm$ 0.03 & 24.1 $\pm$ 0.6 & 7.27 $\pm$ 0.05           &  9.95 $\pm$ 0.20           &  0.26 $\pm$ 0.03           & -0.99 $\pm$ 0.05$^{\rm e}$ \\
   NGC 2146 & SBabp                       & 17.2  & 3.01 $\pm$ 0.001  & 44.71 $\pm$ 0.03 & 37.4 $\pm$ 1.2 & 7.36 $\pm$ 0.08           & 10.30 $\pm$ 0.13           &  1.02 $\pm$ 0.03           &  0.97 $\pm$ 0.03$^{\rm f}$ \\
   NGC 0598 & SAcd                        &  0.84 & 0.38 $\pm$ 0.002  & 42.92 $\pm$ 0.05 & 23.0 $\pm$ 0.7 & 6.68 $\pm$ 0.05           &  8.86 $\pm$ 0.10           & -0.77 $\pm$ 0.05           & -0.80 $\pm$ 0.10 \\ 
   \hline
  \end{tabular}
 \begin{list}{}{}
    \setlength{\itemsep}{0pt}
    \item The columns are: galaxy name; 
          morphological type, from Kennicutt et al.\ (2003); 
          redshift-independent distance (see Kennicutt et al., in prep.); 
          dust/stellar flux ratio (Eqn.~\ref{ratio}); 
          TIR luminosity, using Draine \& Li (2007) calibration; 
          dust temperature, estimated from modified blackbody fit to far-IR SED; 
          dust mass, converted from dust temperatures using Li \& Draine (2001) 500$\mu$m mass absorption coefficient; 
          stellar mass, using Zibetti et al.\ (2009) calibration; 
          star formation rate (SFR), from FUV and TIR luminosities, and from H$\alpha$ and 24~$\mu$m luminosities. 
          The mass and SFR errors are underestimates: they include only formal errors from the fluxes, not systematic errors.
          $^{\mathrm a}$ Morphology obtained from Buta et al.\ (2010), and was different than that listed in Kennicutt et al.\ (2003).  $^{\mathrm b}$ $M_\mathrm{dust}$ estimated from MIPS fluxes only, without longer wavelength fluxes.  $^{\mathrm c}$ Stellar mass-to-light ratios estimated from $g-i$ and $i-H$ colors; the others are estimated from $B-V$ and $V-H$ colors (see Zibetti et al.\ 2009).  $^{\mathrm d}$ Far-UV flux extrapolated from longer wavelengths.  $^{\mathrm e}$ $H\alpha$ flux obtained from Kennicutt et al.\ (2008).  $^{\mathrm f}$ $H\alpha$ flux obtained from Marcum et al.\ (2001).
 \end{list}
 \label{table1}
\end{table*}

\section{Results: Correlations with Galaxy Properties}\label{results}


\subsection{Total Infrared Luminosity}\label{TIRlum}

\subsubsection{Estimating $L_\mathrm{TIR}$}

We begin by analyzing the total infrared (TIR) luminosity of the galaxies in 
our sample.  The TIR luminosity is a useful quantity because it can be 
directly inferred from the IR fluxes, and because it can be used as a proxy 
for the obscured star formation as well as the temperature of dust grains 
(e.g., Dale \& Helou 2002, Draine \& Li 2007).  We follow Draine \& Li 
(2007), and use the 8, 24, 70, 160$\mu$m data from IRAC and MIPS to estimate 
the TIR luminosity:
\begin{equation}
  L_\mathrm{TIR}\,=\,0.95\langle \nu L_\nu\rangle_{7.9} + 1.15\langle\nu L_\nu\rangle_{24} + \langle\nu L_\nu\rangle_{71} + \langle\nu L_\nu\rangle_{160}
 \label{DL07LTIR}
\end{equation}
We have also tested the Dale \& Helou (2002) formula, which uses only the 
MIPS bands, and have obtained very similar results. 
(We have chosen not to include the SPIRE bands in Eqn.~\ref{DL07LTIR}, because a calibration of $L_\mathrm{TIR}$ with these bands has not yet been developed and tested.)

Estimates of $L_\mathrm{TIR}$ are designed to encompass all of the emission from 
PAH particles, very small grains, and large grains, the proportions of which  
depend on the starlight density distribution and the relative abundances 
of the grain populations (Draine \& Li 2007; Compi\'{e}gne et al.\ 2011). 
Therefore, it is useful to compare $L_\mathrm{TIR}$ and the dust/stellar flux ratio, 
in order to analyze the emission from dust grains vis-\'{a}-vis stellar emission.

\subsubsection{Results}

\begin{figure*}
 \includegraphics[width=0.497\hsize]{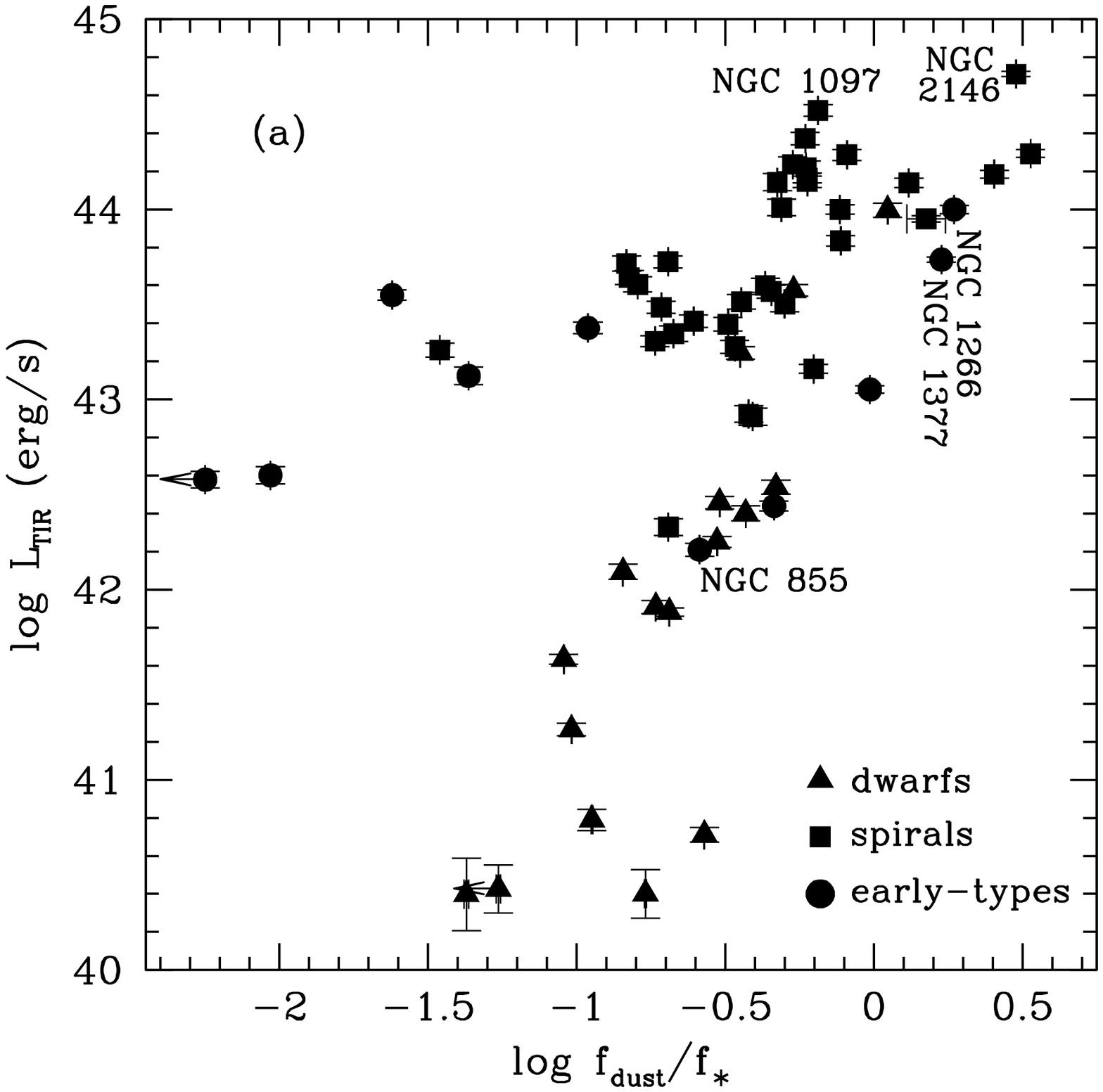} 
 \includegraphics[width=0.497\hsize]{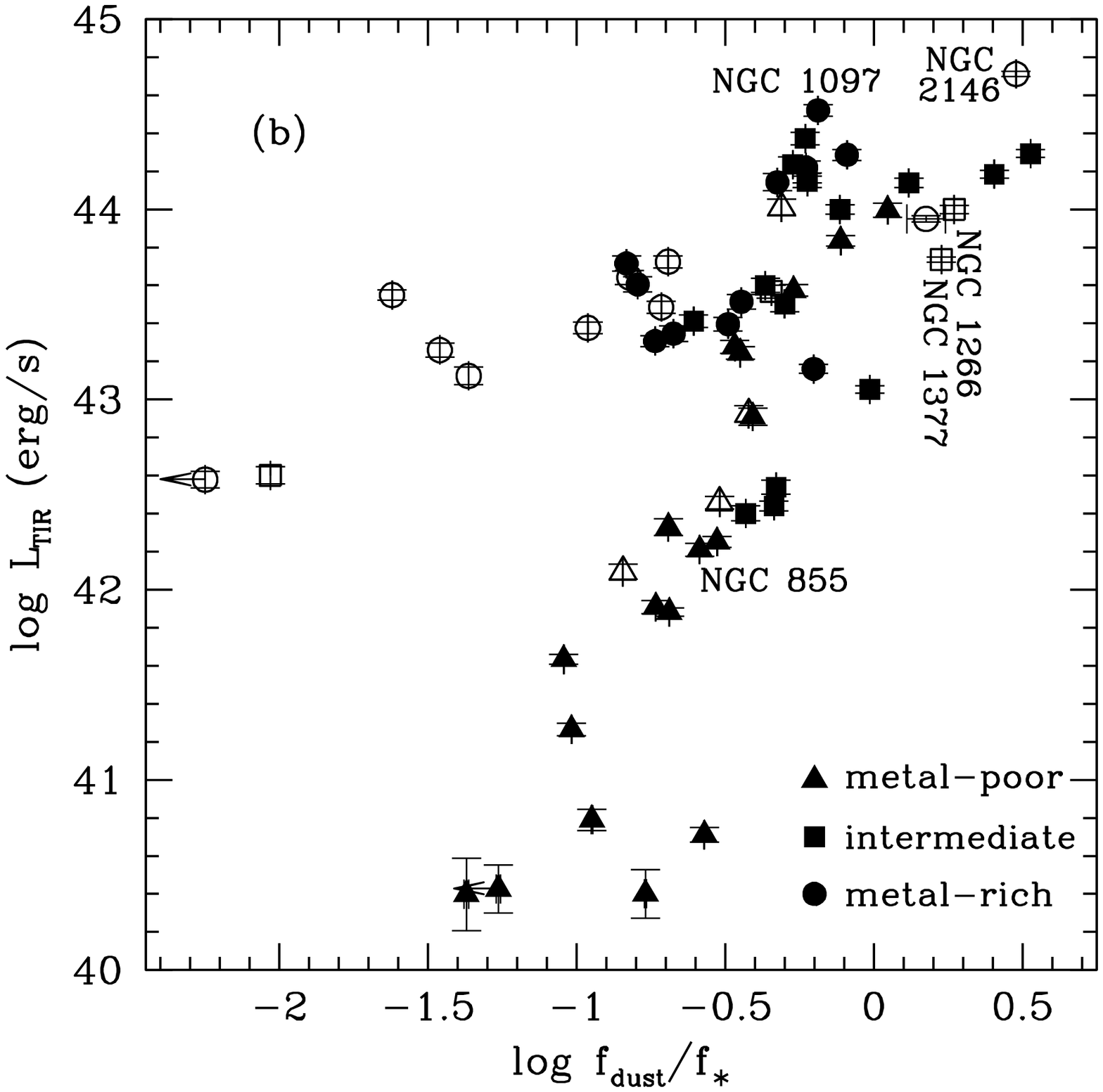} 
 \caption{TIR luminosity estimated using Draine \& Li (2007) formula (using 
          8, 24, 70, 160$\mu$m data) versus dust/stellar emission.
          Left plot: red circles, green squares, and blue triangles have morphology classifications 
          of E and S0, Sa to Scd, and Sd and later type, 
          respectively.
          Right plot: red circles, green squares, and blue triangles have the 
          highest, intermediate, and lowest oxygen abundances (see Section~\ref{metal}); 
          open points indicate abundances estimated from the $B$-band luminosity-metallicity relation (Moustakas et al.\ 2010).
         }
 \label{TIRcorr}
\end{figure*}

The correlation between $L_\mathrm{TIR}$ and the dust/stellar flux ratio of 
the KINGFISH galaxies is shown in Figure~\ref{TIRcorr}.  
As mentioned in Section~\ref{dusttostellar}, the total IR luminosity is very closely related to the quantity $f_\mathrm{dust}$. 
Plotting $f_\mathrm{dust}\times D^2$ versus $f_\mathrm{dust}/f_\ast$ yields a result very similar 
to that shown in the figure.

In Figure~\ref{TIRcorr}a, the galaxies are also labeled by their morphologies, such that we 
distinguish galaxies that are classified as E and S0, Sa to Scd, and Sd and later-type. 
The spiral galaxies NGC~2146 and NGC~1097 have the largest TIR luminosities, while 
the dwarf galaxies DDO~154, DDO~165, and M81~Dwarf~B have the faintest luminosities. 
The late-type spiral galaxies NGC~598 and NGC~7793 have almost exactly the same $f_\mathrm{dust}/f_\ast$ 
and $L_\mathrm{TIR}$; as noted by Smith et al.\ (1984), these galaxies have similar 
photometric and kinematic properties, but different spiral arm structures.

Many of the spirals tend to be found in the locus of $f_\mathrm{dust}/f_\ast$ 
just below unity and $L_\mathrm{TIR}>10^{43}\,\mathrm{erg}\,\mathrm{s}^{-1}$ in the figure. 
Most of the galaxies with small dust/stellar flux ratios and large TIR luminosities 
are early-types, although this may be partly due to a selection effect, as these 
galaxies were selected to be detectable in the IR (Kennicutt et al.\ 2003). 
In contrast, those with small $f_\mathrm{dust}/f_\ast$ 
and small $L_\mathrm{TIR}$ are dwarfs and irregulars. 
Here and in terms of other galaxy properties, some of the earlier-type spirals follow the trends of the E and S0 galaxies, while others have properties more similar to other spiral galaxies. 
The trend in Figure~\ref{TIRcorr} is more complicated than a simple morphological 
distinction, however, as some early-types have relatively large dust/stellar flux 
ratios ($f_\mathrm{dust}/f_\ast\geq1$); as we will show later, these galaxies 
also tend to have larger specific star formation rates and dust temperatures.

Note that Dale et al.\ (2009) have a somewhat similar plot (their Fig. 3), in 
which $L_\mathrm{TIR}/L_B$ is plotted on the horizontal axis, rather than 
the dust/stellar flux ratio.  Their sample is obtained from the 
LVL survey, which was designed to be nearly volume-limited.  As a result, 
it is dominated by faint dwarf galaxies, and some of these fill part of the 
parameter space in the lower right of the figure, with low $L_\mathrm{TIR}$ 
and slightly higher dust/stellar flux ratios than the dwarf galaxies in KINGFISH.
In comparison, KINGFISH is incomplete and has relatively few dwarf galaxies, so the survey 
mostly covers the high-TIR part of the parameter space. 
Lastly, Soifer et al.\ (1989) also performed a similar analysis, plotting 
infrared/visible flux versus FIR luminosity for a sample of IR-bright galaxies in IRAS. 
They obtained a single trend albeit with significant dispersion, but most of 
the dwarf galaxies in KINGFISH would likely not have met their selection criteria.

We summarize the morphology dependence of TIR luminosity and other galaxy 
properties analyzed in this section in Table~\ref{table2}, which 
lists the means and standard deviations of these properties as a function of Hubble type. 
These could be useful as a local benchmark for comparisons with high-redshift studies, 
such as studies of submillimeter galaxies (e.g., Santini et al.\ 2010).
\begin{table*}
 \caption{Mean Galaxy Properties as a Function of Morphological Type}
 \centering
 \begin{tabular}{ l | c c c c c c }
   \hline
    & $\mathrm{log}~f_\mathrm{dust}/f_\ast$ & $\mathrm{log}~M_\mathrm{dust}/M_\ast$ & $\mathrm{log}~L_\mathrm{TIR}$ & $T_\mathrm{dust}$ & $\mathrm{log}~SSFR_{H_\alpha+24\mu{\rm m}}$ & $\mathrm{log}~SSFR_\mathrm{FUV+TIR}$ \\
    & & & (log~erg~$s^{-1}$) & (K) & (log~$M_\odot~\mathrm{yr}^{-1}$) & (log~$M_\odot~\mathrm{yr}^{-1}$) \\
   \hline
   All Galaxies & $-0.52\pm0.07\pm0.54$ & $-2.95\pm0.09\pm0.68$ & $43.12\pm0.14\pm1.08$ & $27.0\pm0.6\pm4.3$ & $-9.84\pm0.10\pm0.71$  & $-9.74\pm0.09\pm0.74$  \\
   Dwarfs       & $-0.69\pm0.09\pm0.36$ & $-2.70\pm0.13\pm0.53$ & $41.88\pm0.26\pm1.08$ & $27.6\pm0.9\pm3.8$ & $-9.52\pm0.12\pm0.43$  & $-9.27\pm0.11\pm0.47$  \\
   Spirals      & $-0.35\pm0.07\pm0.39$ & $-2.83\pm0.08\pm0.49$ & $43.74\pm0.09\pm0.51$ & $25.8\pm0.6\pm3.5$ & $-9.90\pm0.12\pm0.65$  & $-9.76\pm0.09\pm0.53$  \\
   Early-types  & $-0.87\pm0.28\pm0.88$ & $-3.77\pm0.26\pm0.83$ & $43.07\pm0.18\pm0.57$ & $30.2\pm2.0\pm6.0$ & $-10.15\pm0.37\pm1.04$ & $-10.45\pm0.34\pm1.06$ \\
   \hline
  \end{tabular}
 \begin{list}{}{}
    \setlength{\itemsep}{0pt}
    \item KINGFISH galaxy properties: mean$\pm$bootstrap error$\pm$standard deviation. 
          There are 17 
          dwarf and irregular galaxies (Sd and later), 
          35 spirals (Sa to Scd), 
          and 10 early-types (E and S0). 
          $f_\mathrm{dust}/f_\ast$, $M_\mathrm{dust}/M_\ast$, $L_\mathrm{TIR}$, 
          $T_\mathrm{dust}$, and $SFR$ are discussed in Sections~\ref{dusttostellar}, 
          \ref{masses}, \ref{TIRlum}, \ref{FIRcolors}, and \ref{SFRsec}, respectively. 
          $SSFR$ refers to specific star formation rate, $SFR/M_\ast$; the quantities in these 
          two columns include galaxies with $H\alpha$ fluxes from Kennicutt et al.\ (2008) 
          and galaxies with extrapolated FUV fluxes (see Section~\ref{dusttostellar}). 
 \end{list}
 \label{table2}
\end{table*}

As mentioned above, approximately half of the KINGFISH galaxies have strong 
bars, and it is possible that the presence of a bar affects emission by dust and 
stars in the central region of a galaxy.  
We find that the mean of the log dust/stellar flux ratio is 
$-0.57\pm0.10$ for weakly barred galaxies (no bar or SA) and $-0.48\pm0.09$ 
for strongly barred galaxies (SAB or SB). 
On the other hand, for the total infrared luminosity, the means 
are $\overline{\mathrm{log}\,L_\mathrm{TIR}}=42.9\pm0.2$ and $43.3\pm0.2$ 
for weakly and strongly barred galaxies, respectively.  
This is consistent with the observation that most IR-selected starburst galaxies 
are barred (e.g., Hunt \& Malkan 1999). 
Nevertheless, the bar fraction has also been observed to be higher among galaxies with higher 
optical luminosities, 
because these galaxies became dynamically cool and sufficiently massive to host bars earlier than fainter galaxies (Sheth et al.\ 2008). 
Thus one might expect barred 
galaxies to have brighter stellar and infrared luminosities than weakly barred ones. 
Our data in fact bear this out: the mean stellar and dust luminosities, 
which we quantify as $f_\ast(4\pi D^2)$ and $f_\mathrm{dust}(4\pi D^2)$, are both higher 
in barred galaxies by $\approx0.43$~dex, the same factor as $\mathrm{log}\,L_\mathrm{TIR}$. 
Therefore, it is simply the case that strongly barred galaxies in KINGFISH are 
more luminous than weakly barred ones, and this effect cancels in the dust/stellar flux ratios.

In Figure~\ref{TIRcorr}b, we show the metallicity 
dependence of the relation between the TIR luminosity and dust/stellar flux 
ratio.  Firstly, it can be seen that the metallicity dependence and 
morphology dependence are similar: dwarf galaxies tend to be relatively 
metal-poor, spiral galaxies tend to have intermediate metallicities, and 
early-type galaxies tend to be metal-rich.  Nonetheless, it is not a 
one-to-one relationship: some early-type spirals with large $L_\mathrm{TIR}$ 
are also metal-rich (such as NGC~1097), while some intermediate-metallicity galaxies are 
early-types with relatively large $f_\mathrm{dust}/f_\ast$ (such as NGC~855). 


In general, from the morphological and metallicity dependence of 
$L_\mathrm{TIR}$ and $f_\mathrm{dust}/f_\ast$, 
we can tentatively infer an evolutionary sequence from the figure, 
such that as a typical late-type galaxy grows and becomes more luminous, it 
becomes more metal enriched and has a larger dust fraction; this is also 
accompanied by more emission by stars, a larger metallicity, and a growing 
stellar mass (either by star formation or a merger), as the galaxy becomes an early-type. 
For the early-types, either stellar mass growth outweighs dust production, 
because the dense molecular clouds in which dust is typically accreted have dissolved, 
or a substantial amount of dust grains are ejected or destroyed, such as by supernovae shockwaves and thermal `sputtering' (e.g., Draine \& Salpeter 1979; Dwek 1998; Pipino et al.\ 2011). 
This interpretation of Figure~\ref{TIRcorr} is merely speculative, however, 
and assumes that typical present-day late-types resemble past stages of present-day early-types; 
that is, it assumes that the growth of galaxy disks precedes that of bulges (e.g., Bournaud et al.\ 2009; Ceverino et al.\ 2010), 
although there is evidence that this assumption may be too simplistic (e.g., MacArthur et al.\ 2009; Bundy et al.\ 2010). 

In contrast, the dwarf and irregular galaxies have lower TIR luminosities (most have $L_\mathrm{TIR}<10^{42.5}\mathrm{erg}~s^{-1}$) and relatively little 
emission from dust (usually $f_\mathrm{dust}/f_\ast<0.3$).  
They appear to be a distinctly different galaxy 
population, exhibiting different properties than typical late-type spirals and 
inhabiting different environments (e.g., Leroy et al.\ 2008; Gavazzi et al.\ 2010). 

In any case, although general trends are apparent in Figure~\ref{TIRcorr}, there are plenty 
of exceptions and variation within the KINGFISH sample. 
Note that the sample was selected not to have strong AGN, defined as an AGN that dominates substantial portions of a galaxy's spectrum, so it is unlikely that AGN contribute much to this variation.

\subsection{Dust and Stellar Mass}\label{masses}

Next, in order to add to this picture of galaxy evolution, we can use the infrared 
SEDs of the galaxies in our sample to estimate dust and stellar masses. 
These are more physical quantities than fluxes, but they require some model 
assumptions, and hence have additional systematic uncertainties.

\subsubsection{Estimating $T_\mathrm{dust}$ and $M_\mathrm{dust}$}\label{TdMd}

We estimate dust masses using dust temperatures, which are determined from 
simple fits to the FIR SED (e.g., Hildebrand 1983).  We use the MIPS and SPIRE FIR and submm 
($70-500\mu$m) flux densities and perform fits using a single temperature 
blackbody modified by an emissivity law proportional to $\lambda^{-\beta}$.  
Our approach is similar to that of Engelbracht et al.\ (2010) and Gordon et al.\ (2010), who also applied this method to galaxies with MIPS and SPIRE data. 
The use of a single temperature fit for whole galaxies essentially yields an average 
dust temperature; in practice, most galaxies have multiple dust components at 
different temperatures, such as in photo-dissociation regions and in the diffuse ISM, 
although one component might dominate. 
We assume $\beta=1.5$, which is consistent with recent observational constraints at far-IR and submm wavelengths (e.g., Dunne \& Eales 2001; Paradis et al.\ 2009; Gordon et al.\ 2010). 
In any case, the resulting dust temperatures are slightly dependent upon the assumed emissivity, and it is in a systematic way, such that $\beta=2$ would yield slightly lower temperatures by a few degrees for the whole sample (e.g., Bendo et al.\ 2003). 
We also do not attempt to account for the `submm excess' inferred by some authors 
which may be due to a very cold dust component or to a wavelength dependent 
emissivity law (e.g., Galametz et al.\ 2009; Gordon et al.\ 2010). 
We infer the uncertainties of 
the temperatures with a Monte Carlo analysis that includes the flux errors.  

Most of the resulting temperatures range from $20-35$ K, with NGC~1512 
having the coldest temperature ($\approx21\,$K), and NGC~1377 and NGC~2146 having 
exceptionally warm ones ($\approx43\,$K and $37\,$K, respectively). 
NGC~1404 yields an extremely cold temperature, 
but its FIR and submm flux densities are very small, and the 70 and 160 $\mu$m 
fluxes quoted in Dale et al.\ (2007) appear to be due to a background source, 
so we discard the galaxy's temperature as unrealistically low. 
For most other galaxies, a modified blackbody provides a good fit to the FIR SED. 
%
The temperatures and uncertainties are listed in Table~\ref{table1}.


Given $T_\mathrm{dust}$ of a galaxy, we estimate the dust mass with 
the following:
\begin{equation}
  M_\mathrm{dust} \,=\, \frac{f_\lambda (4\pi D^2)}
                             {\kappa_{\mathrm{abs},\lambda}\,(4\pi B_\lambda(T_\mathrm{dust}))}
 \label{Mdust}
\end{equation}
where $f_\lambda$ is the flux density, $D$ is the distance from the galaxy, $B_\lambda$ is 
the Planck function, which is $2ckT/\lambda^4$ in the Rayleigh-Jeans limit (and the additional $4\pi$ factor is due to integrating over steradians). 

$\kappa_{\mathrm{abs},\lambda}$ is the mass absorption coefficient, 
which we take from Draine (2003). 
The Draine (2003) model assumes that the dust consists of a mixture of carbonaceous grains 
and amorphous silicate grains, 
with dust grain size distributions consistent with the wavelength-dependent 
extinction in the local Milky Way, with $R_V=3.1$ (Weingartner \& Draine 2001). 
This model also assumes that most of the dust is heated by the diffuse 
radiation field, while the rest is heated by luminous stars with intense starlight.  

We compute the masses at $\lambda=500\mu$m, 
in order to minimize the dependence on the temperature, although 
the uncertainties of the flux densities are larger than at shorter wavelengths. 
At $500\mu$m, the mass absorption coefficient is 
$\kappa_\mathrm{abs}=0.95\,\mathrm{cm}^2/\mathrm{g}$ 
(which is $\approx20\%$ lower than the value quoted in Li \& Draine (2001)). 
Importantly, note that the absorption coefficients in these models have approximately 
$\kappa \propto \lambda^{-2}$, while many of the galaxies have an emissivity 
closer to $\beta\approx-1.5$, which entails a wavelength dependence of the estimated dust masses. 
Masses estimated at $\lambda=250\mu$m and $350\mu$m are lower by 
$\approx0.5$~dex and 0.25~dex, respectively.


For the galaxies lacking SPIRE fluxes, NGC~1404 and DDO~154, we estimate dust masses using 
the procedure outlined in Draine \& Li (2007, Section 9.5), using the 
galaxies' fluxes at 8, 24, 70, and 160 $\mu$m.  The Draine \& Li (2007) model 
is an updated version of the one developed by Li \& Draine (2001). 
These dust masses are less reliable than the ones 
estimated from dust temperatures 
because the submm SEDs are less strongly constrained. 
The quoted errors of all of the dust masses consist only of the formal 
errors, and do not include systematic uncertainties. 

The dust masses of the KINGFISH galaxies fall between 
$10^4-10^8\,M_\odot$.  The dwarf and irregular galaxies have the lowest masses, 
while early-type spirals such as NGC~1097 and NGC~7331 have the highest masses.
Consistent with Masters et al.\ (2010a), luminous disk-dominated spirals such 
as NGC~4254, NGC~5457, and NGC~6946 have relatively large dust masses ($M_\mathrm{dust}\approx10^{7.5}\,M_\odot$). 
The dust masses and uncertainties are listed in Table~\ref{table1}.

Dust mass is strongly correlated with total infrared luminosity, which we 
analyzed in the previous section.  The distribution of the dust/stellar 
flux ratio as a function of dust mass is qualitatively similar to Figure~\ref{TIRcorr}, 
but with larger uncertainties.

Fifteen of our galaxies are included in the subsample of Draine et al.\ (2007) 
of galaxies that have SCUBA data. 
We have compared our dust masses to theirs, and our masses are systematically 
lower by $\approx$0.2-0.4 dex. 
Some of their distances are shorter, and since distance appears in quadrature 
in Eqn.~\ref{Mdust}, accounting for this makes the discrepancy slightly larger.  
The discrepancy may be partly due to our single-temperature fit including the 
70~$\mu$m flux, which may have a contribution of small dust grains stochastically 
heated by starlight (Draine \& Li 2007); galaxies with slightly larger temperature estimates would then have slightly smaller masses. 
The wavelength dependence of dust emissivity, mentioned above (and see Paradis et al.\ 2009), 
could also be an important factor. 
Two galaxies with large discrepancies are 
NGC~5713, for which Draine et al.\ (2007) obtained a mass nearly 0.9 dex larger and for which a LMC dust model was favored, and NGC~4631, for which they obtained a mass 0.8 dex larger although the result depends on the allowed range of starlight intensity. 
Our dust mass for NGC~3077 is slightly larger than that estimated by Walter et al.\ (2011), which 
we believe is mostly due to their smaller aperture size. 
A detailed analysis of the dust masses and other dust properties of KINGFISH galaxies 
will be presented in Dale et al.\ (in prep.) and Aniano et al.\ (in prep.). 

\begin{figure*}
  \includegraphics[width=0.497\hsize]{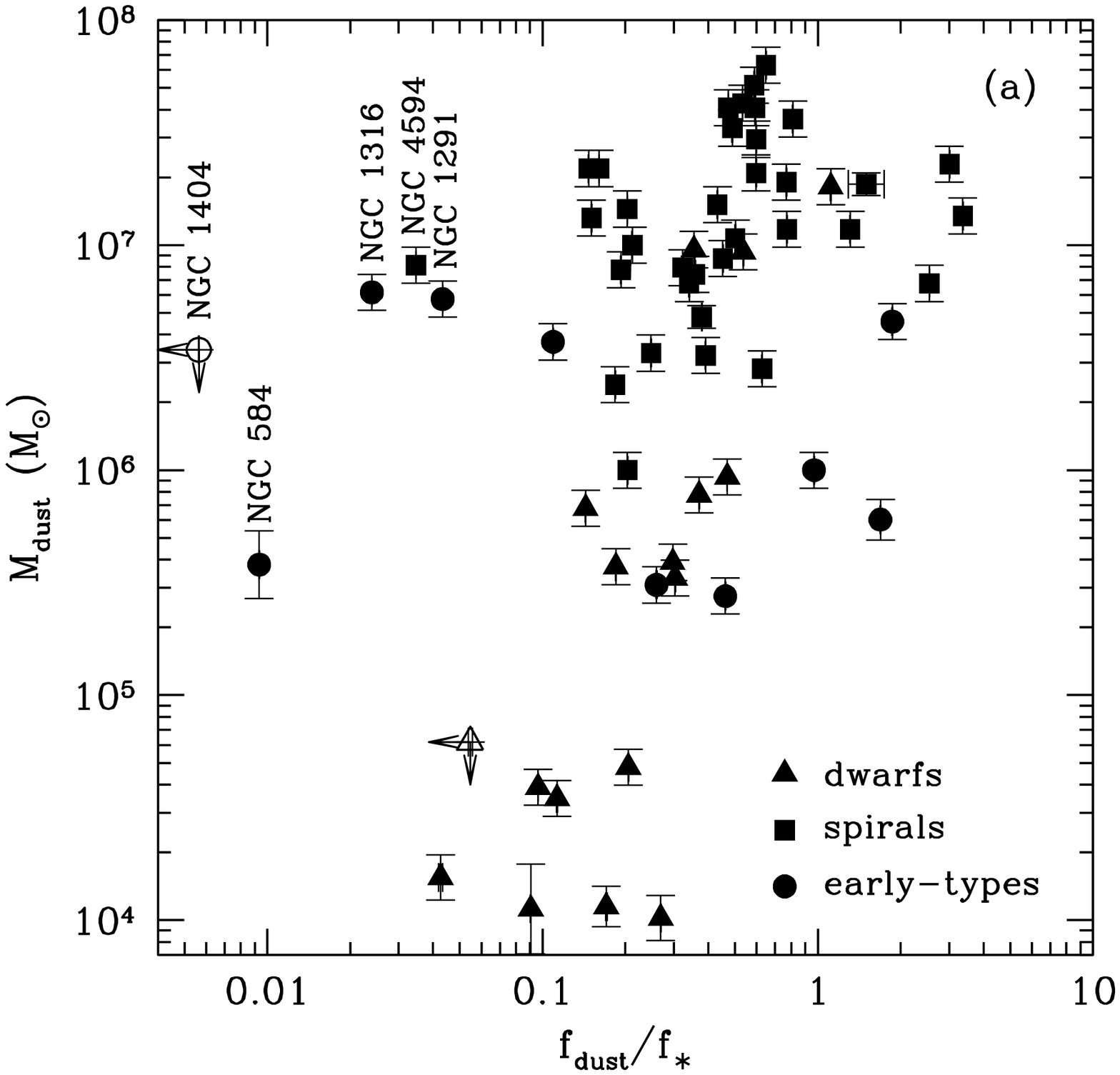}
  \includegraphics[width=0.497\hsize]{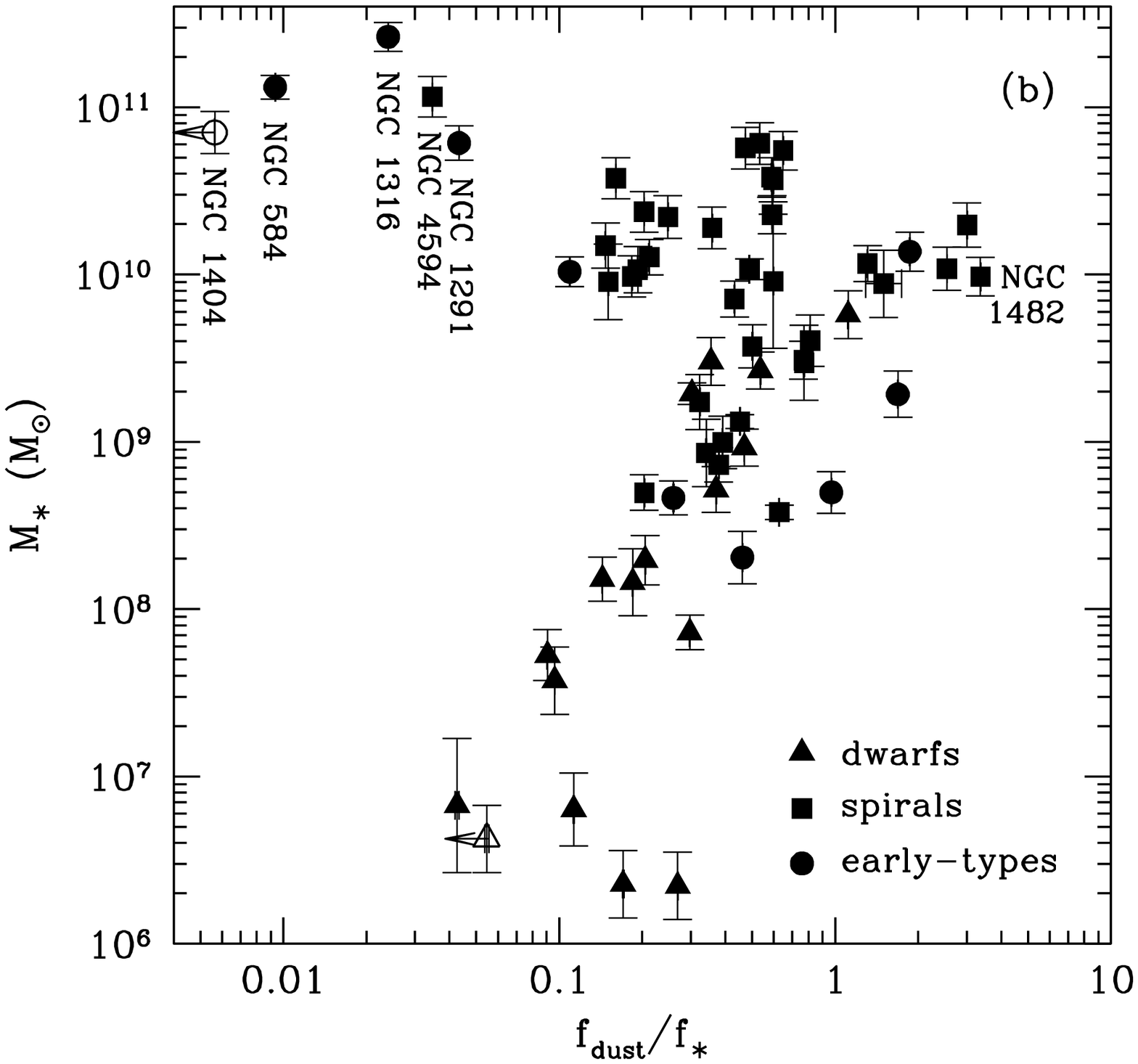}
  \caption{
           Dust mass (left) and stellar mass (right) versus dust/stellar flux ratio (Eqn.~\ref{ratio}). 
           Red circles, green squares, and blue triangles indicate galaxies with E and S0,
           Sa to Scd, and Sd and later-type morphologies, respectively.  Open points indicate
           galaxies for which we do not have detected SPIRE fluxes.
          }
  \label{fratiovsM}
\end{figure*}

\subsubsection{Estimating $M_\ast$}\label{mstar}

We estimate the stellar masses from Zibetti et al.\ (2009), using optical and 
near-IR colors with $H$-band luminosity.  In particular, they combine 
stellar population synthesis (SPS) models with simple prescriptions for dust 
attenuation.  They use an updated version of the Bruzual \& Charlot (2003) 
SPS models, which include revised prescriptions for the thermally pulsing 
asymptotic giant branch (TP-AGB) evolutionary phase, with a two component 
star formation history, consisting of a continuous, exponentially declining 
mode with random bursts superimposed.  The Zibetti et al.\ (2009) model 
outputs stellar mass-to-light ratios $\mathrm{log}\,M_\ast/L_H(B-V,V-H)$, and for the LVL 
galaxies, which have $ugriz$-band SDSS data, we use $\mathrm{log}\,M_\ast/L_H(g-i,i-H)$. 
In other words, the stellar masses are calculated with the following: 
\begin{equation}
  \mathrm{log}\,M_\ast \,=\, \mathrm{log}\,L_H \,+\, f(c_\mathrm{opt},c_\mathrm{optNIR}) ,
 \label{Mstar}
\end{equation}
where $c_\mathrm{opt}$ is the optical color ($B-V$ or $g-i$) and 
$c_\mathrm{optNIR}$ is the optical-NIR color ($V-H$ or $i-H$). 
As a function of two colors, the model's mass-to-light ratios typically have 
0.1-0.2 dex scatter. 
Before computing the masses, we have corrected the colors for foreground 
Galactic extinction, using values taken from Schlegel et al.\ (1998).
We assume a universal Kroupa (2001) initial mass function (IMF). 
The assumed IMF affects the inferred masses and SFRs systematically 
(e.g., for a Salpeter IMF, they would be higher by a factor of 1.8 (Borch et al.\ 2006) and 1.5 (Calzetti et al.\ 2010), respectively), 
while the relative distributions are robust.


Most of the resulting stellar masses are in the range $10^7-10^{11}\,M_\odot$. 
The S0 galaxy NGC~1316, whose SED is shown in Figure~\ref{SEDs}, has the largest stellar mass. 
Like the dust masses, the irregular galaxies have the lowest stellar masses; however, NGC~3077 is an 
exception: it is a relatively massive dwarf galaxy in the M81 group, whose starburst activity may have been triggered by interactions with its neighbors (Walter et al.\ 2002). 
Note that other models (e.g., Bell et al.\ 2003; Sani et al.\ 2011) yield relatively similar but not identical stellar masses for these galaxies; we refer the reader to Zibetti et al.\ (2009) for a comparison with and discussion of other models.


\subsubsection{Results}

First, we show in Figure~\ref{fratiovsM} 
the dust mass and stellar mass as a function of the dust/stellar flux ratio. 
The two figures are similar, because 
the galaxies in this sample have such strongly correlated dust and stellar masses 
(unlike high-redshift submm galaxies, for example, which have a more scattered correlation; Santini et al.\ 2010). 

The galaxies are marked by their morphology classes in Figure~\ref{fratiovsM}. 
Four early-types (NGC~1404, NGC~584, NGC~1316, NGC~1291) 
and an earlier-type spiral (NGC~4594) are outliers, with low $f_\mathrm{dust}/f_\ast$ and large dust and stellar masses. 
It is also interesting that $f_\mathrm{dust}/f_\ast$ is correlated with 
the dust and stellar masses for the dwarf/irregular and late-type spiral galaxies, with 
Spearman rank correlation coefficients\footnote{The Spearman rank correlation 
coefficient may have a value between $-1$ and 1.  A positive (negative) value 
indicates an (anti)correlation, and a value of 0 indicates no correlation.} 
of $r_s=0.52$ and $0.51$ (which implies approximately 95\% significance), respectively. 
Most dwarf galaxies lack substantial dust emission (e.g., Walter et al.\ 2007), but the few of them with $f_\mathrm{dust}/f_\ast \geq0.3$ (mostly Sd galaxies and NGC~3077) have relatively large dust and stellar masses. 

In both figures, but especially in the plot showing the stellar mass dependence, 
(Fig.~\ref{fratiovsM}b), there appears to be a transition between two populations of galaxies.  
In particular, there appears to be a ``transition'' stellar mass at $M_\ast\sim10^{10}\,M_\odot$, 
such that less massive galaxies follow a steep relation with dust/stellar flux, while more 
massive galaxies occur on a flat or slightly negative relation.  
We will later show that a similar transition appears to occur between specific 
star formation rate and dust/stellar flux (see especially Fig.~\ref{SSFRcorr}b in 
Section~\ref{SFRsec}).  This mass scale of $10^{10}\,M_\odot$ is similar to the transition 
stellar mass determined by Kauffmann et al.\ (2003; see also Schiminovich et al.\ 2007), 
above which galaxies have high stellar mass surface densities, high concentration indices typical 
of bulges, old stellar populations, and low SFRs and gas masses.  Intriguingly, the maximum 
$f_\mathrm{dust}/f_\ast$ occurs in the transition region, and perhaps NGC~1482 is a 
candidate for a transition galaxy in this context; however, note that some other galaxies in 
this region are peculiar, and in some cases are interacting with a neighbor. 
In any case, it is interesting that the transition in $f_\mathrm{dust}/f_\ast$ occurs at a similar stellar mass as the previously observed transition in $D_n(4000)$ (4000-\AA  
break strength, quantifying the star formation history), $\mu_\ast$ (stellar mass surface density), and specific SFR.

\begin{figure*}
  \includegraphics[width=0.497\hsize]{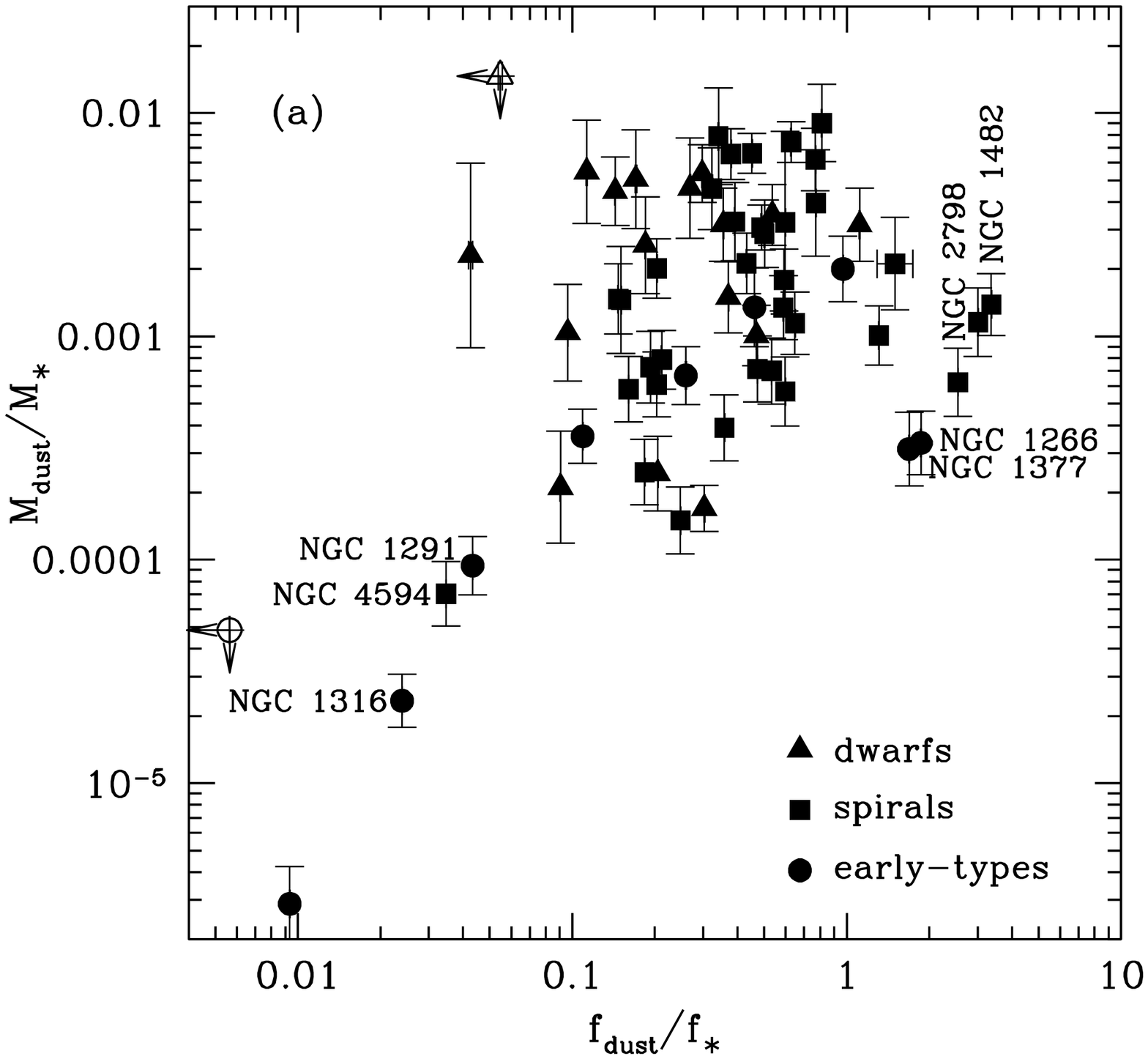} 
  \includegraphics[width=0.497\hsize]{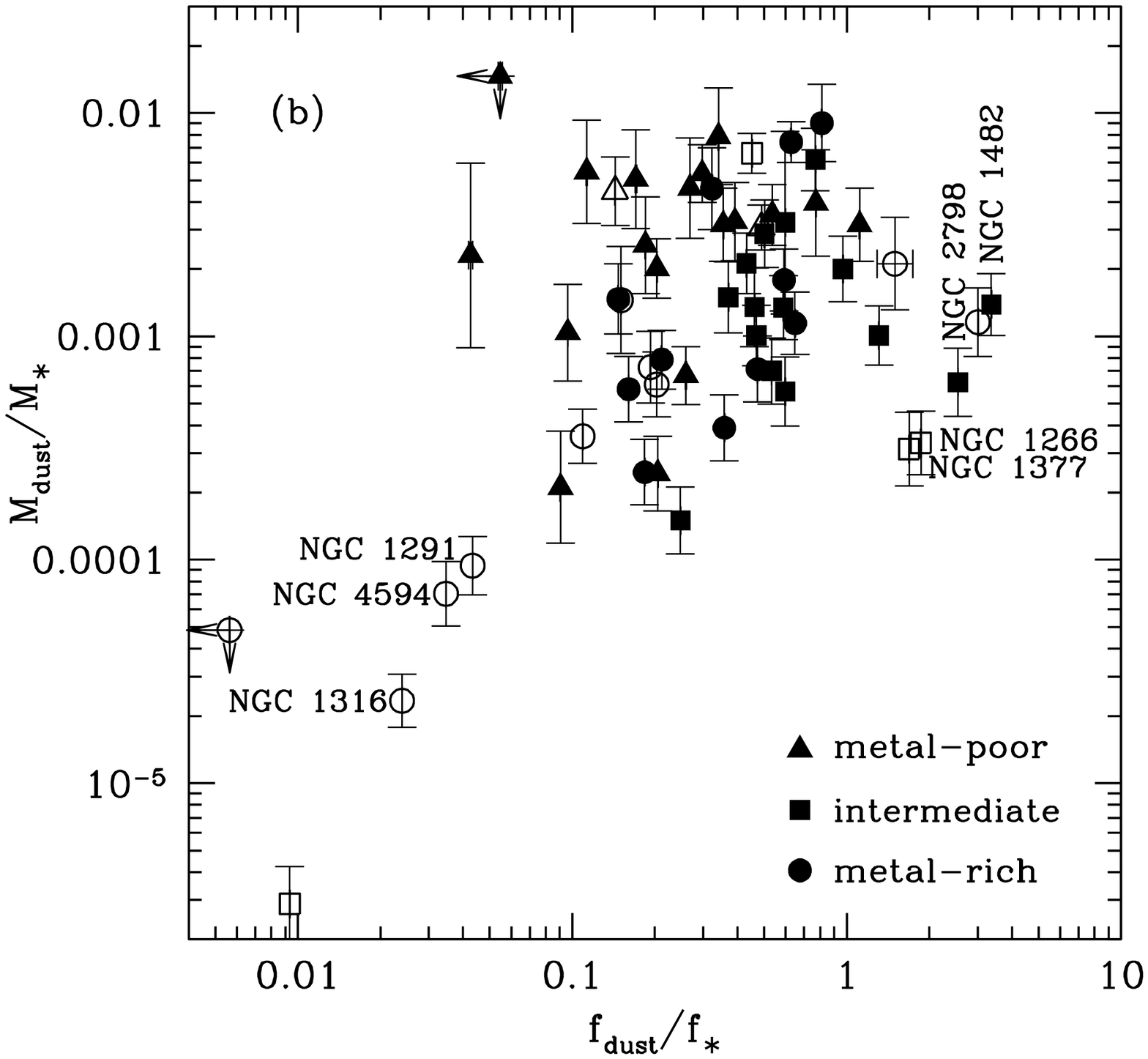} 
  \caption{Dust/stellar \textit{mass} ratio versus dust/stellar \textit{flux} ratio.
           Left plot: red circles, green squares, and blue triangles indicate galaxies with E and  
           S0, Sa to Scd, and Sd and later-type morphologies, respectively.  Open points indicate 
           galaxies for which we do not have detected SPIRE fluxes and dust temperatures; 
           these dust masses may have additional systematic uncertainties.
           Right plot: red circles, green squares, and blue triangles have the 
           highest, intermediate, and lowest oxygen abundances, from Moustakas et al.\ (2010); 
           open points indicate abundances estimated from the luminosity-metallicity relation.
          }
  \label{Mratios}
\end{figure*}

Next, we compare the ratio of dust/stellar mass to the dust/stellar 
flux ratios in Figure~\ref{Mratios}.  Note that the $M_\mathrm{dust}/M_\ast$ 
distribution is different than that of $f_\mathrm{dust}/f_\ast$: 
the $f_\mathrm{dust}/f_\ast$ distribution is centrally peaked with small 
and large outliers (see Figure~\ref{fracdist}), while the majority of the 
$M_\mathrm{dust}/M_\ast$ ratios are between $10^{-4}$ and $10^{-2}$, 
with the aforementioned early-types and NGC~4594 having lower values. 
{NGC~584} in particular has a stellar mass that is more than five orders of magnitude larger than its dust mass. 

As stated in Table~\ref{table2}, the mean log dust/stellar mass ratio is $-2.95\pm0.09$ for the sample, which is smaller than the low-redshift value given by Dunne et al.\ (2011), but the results appear to be consistent given the different sample selections. 
Dunne et al.\ find that the dust/stellar mass ratio depends on metal enrichment, which we also discuss below, and increases with increasing redshift (see also Bussmann et al.\ 2009).

For the general galaxy population in KINGFISH, there is only a weak correlation between 
$M_\mathrm{dust}/M_\ast$ and $f_\mathrm{dust}/f_\ast$ (Spearman rank 
$r_s=0.30$, less than 95\% significance), with substantial scatter.  
Some of the scatter is likely due to systematic uncertainties in the dust and stellar masses, 
as well as to the errors of the $500\mu$m flux densities, which were used to estimate $M_\mathrm{dust}$.  
Spiral and dwarf galaxies appear to populate the same locus on the plot.  
DDO~154 and DDO~165, both dwarf galaxies, are exceptions, with particularly small stellar masses. 
Three Sa galaxies---NGC~4594 
in the lower left of the plot and NGC~1482 and NGC~2798 at the high 
$f_\mathrm{dust}/f_\ast$ end---are also outside of the central locus of the plot, and appear 
to follow the trend of early-type galaxies. 

More clearly than the rest of the sample, the early-types exhibit a strong 
correlation between the dust/stellar ratios (Spearman rank $r_s=0.56$, $95\%$ significance). 
An interesting question to ask 
is why the early-types have such a large range of dust/stellar flux and mass 
ratios, spanning three orders of magnitude. 
One possibility is that these galaxies have a wide range of radiation field intensities 
and star formation rates (see Section~\ref{SFRsec}), 
as well as widely varying gas/stellar ratios (Young et al.\ 2009). 

In particular, all but two of the early-types 
are lenticular galaxies, and many of them have similar masses and TIR 
luminosities.  If we look at one pair of S0s, NGC~1316 and NGC~1291 have 
extremely low dust/stellar ratios, 
and appear to be similar to the elliptical galaxies (NGC~1404 and NGC~584).  
Looking at another pair 
of S0s, NGC~1266 and NGC~1377 have particularly high dust/stellar ratios.  
For NGC~1266, we believe this could be due to an active nucleus (see Dale et 
al. 2007; Smith et al. 2007) which could be heating dust in the central region; however, 
this galaxy is likely an exception, as galaxies in the sample were selected 
not to have luminous AGN (Kennicutt et al.\ 2003). 
For NGC~1377, Roussel et al.\ (2006) argues that it is undergoing an opaque 
nascent burst of star formation, and the intense radiation field is 
significantly heating the dust.  

These two S0's and the Sa's mentioned previously, NGC~1482 and NGC~2798, 
are most offset from the trend of the rest of the sample. 
They are relatively bright in the far-IR, but this is not accompanied by 
a large dust mass.  The offset may be explained by their relatively warm 
dust temperatures, which we discuss in Section~\ref{FIRcolors}. 
In principle, large stellar mass-to-light ratios could also contribute to 
offsets in this direction, but the $M_\ast/L$ of these galaxies are not particularly large.

We emphasize that 
\textit{the outliers and substantial scatter evident in Figure~\ref{Mratios} 
highlight the danger of using $f_\mathrm{dust}/f_\ast$ as a 
proxy for $M_\mathrm{dust}/M_\ast$}.  The two quantities are certainly related, 
but they probe different physical processes, with different dependencies on a galaxy's star formation history and history of dust production and destruction. 
Considering the selection criteria of the KINGFISH survey, only one fourth of which 
is composed of faint dwarf galaxies, the ``true'' scatter between these dust/stellar 
ratios is probably even larger. 
%
For the rest of this paper, we continue to focus on relations between 
$f_\mathrm{dust}/f_\ast$ and other galaxy properties, in order 
to keep the analysis as empirical as possible; subsequent KINGFISH papers 
will provide more detailed analyses of the galaxies' dust and stellar masses. 

We also briefly note that the dust/stellar mass ratios vary slightly 
with bar strength: the mean ratio for strongly barred galaxies (SAB or SB) 
is $\overline{\mathrm{log}\, M_\mathrm{dust}/M_\ast}=-2.84\pm0.58\,$(rms), 
while galaxies with weak or no bars have a slightly lower mean ratio, 
$-3.06\pm0.74\,$(rms). 
This dependence may be related to the fact that strongly barred galaxies 
have warmer dust temperatures in their inner regions, 
likely due to bar-induced star formation in these regions 
(Engelbracht et al.\ 2010). 
Nonetheless, the variation of $M_\mathrm{dust}/M_\ast$ among barred and 
non-barred galaxies is much greater than the difference between them.

The metallicity dependence of the dust/stellar correlation 
(Figure~\ref{Mratios}b) is similar to the morphology dependence, because of 
the relation between morphology and metallicity (e.g., Moustakas et al.\ 2010). 
Nonetheless, there are a 
few differences.  The correlation is strongest for metal-rich galaxies ($r_s=0.80$, $99.9\%$ significance), 
while some galaxies with intermediate metallicities have lower dust/stellar 
mass ratios than one might expect, given their flux ratios. 
These are S0 and Sa galaxies with larger than average star formation rates 
(see Section~\ref{SFRsec}); their new stars could be significantly heating the 
dust and increasing the FIR emission, without significantly increasing the dust mass.

In addition, the fact that the dust/stellar flux ratio is more metallicity-dependent 
than the dust/stellar mass ratio (comparing the metal-poor and intermediate-metallicity 
galaxies) suggests that the scatter between them is partly due to metallicity. 
This appears to be consistent with Groves et al.\ (2008), who found in their SED models 
that the IR peak shifts to longer wavelengths and becomes broader with increasing metallicity, 
due to the increasing dust column and the increasing mechanical luminosity of starburst regions. 
The residuals are also significantly correlated with dust temperature, suggesting 
that the scatter is also partly due to temperature, 
and the geometrical distributions of dust and stars within the galaxies. 
This is not surprising: dust mass is inversely related to dust
temperature (Eqn.~\ref{Mdust}), and as we show in the next section, the dust/stellar flux ratio is positively correlated with temperature, which implies that galaxies with warmer dust 
tend to have more dust emission, and tend to be located towards the lower right of Figure~\ref{Mratios}. 

Lastly, we note that the metallicity dependence of the dust/stellar mass 
ratio here is qualitatively similar to that of the dust/gas mass ratio in 
Draine et al.\ (2007), who studied a similar sample of galaxies. 
For the galaxies with submm data, Draine et al.\ (2007) find that the 
dust/gas ratio gradually increases by up to $0.3$ dex over the  range of 
gas-phase metallicity.  
We also find that our metal-poor galaxies tend to have lower 
dust/stellar flux ratios than intermediate-metallicity galaxies. 
Nevertheless, there 
is no significant metallicity dependence of the dust/stellar \textit{mass} ratios. 
Therefore, the scatter in Figure~\ref{Mratios} may be partly explained by 
the differential dependence on metallicity. 

Furthermore, since lower-metallicity galaxies have lower dust/gas mass ratios 
but not necessarily lower dust/stellar mass ratios, 
it is possible that they have 
lower stellar mass for a given amount of gas---that is, that their star formation is less 
efficient than more metal-rich galaxies.
This is consistent with Lee et al.\ (2006), who found that metallicity is inversely 
related to the gas/stellar mass ratio. 
Note, however, that specifically for massive galaxies, Schiminovich et al.\ (2010) argue that 
star formation efficiency is independent of stellar mass and stellar mass surface density.


\subsection{Far-Infrared Colors}\label{FIRcolors}

\begin{figure*}
 \includegraphics[width=0.497\hsize]{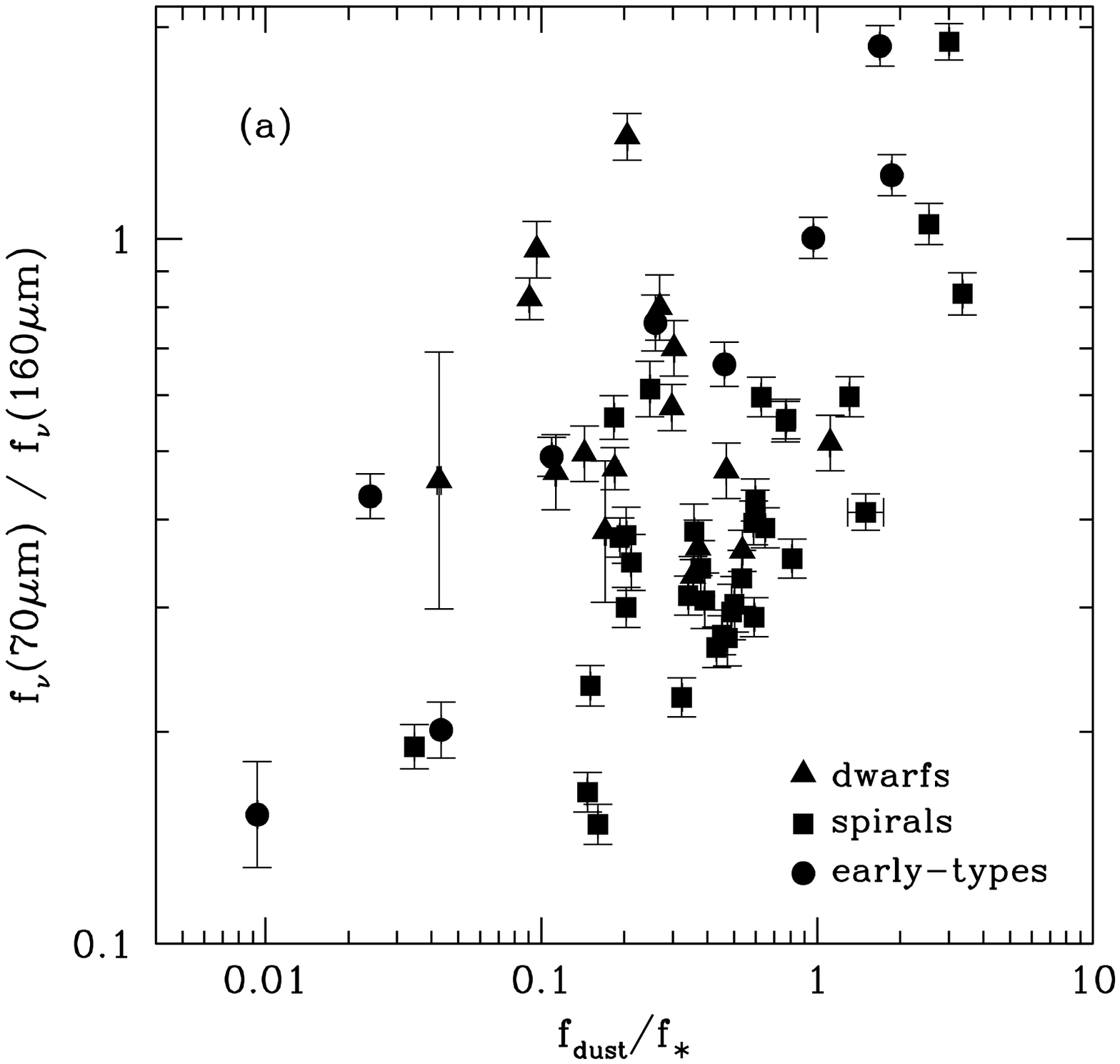} 
 \includegraphics[width=0.497\hsize]{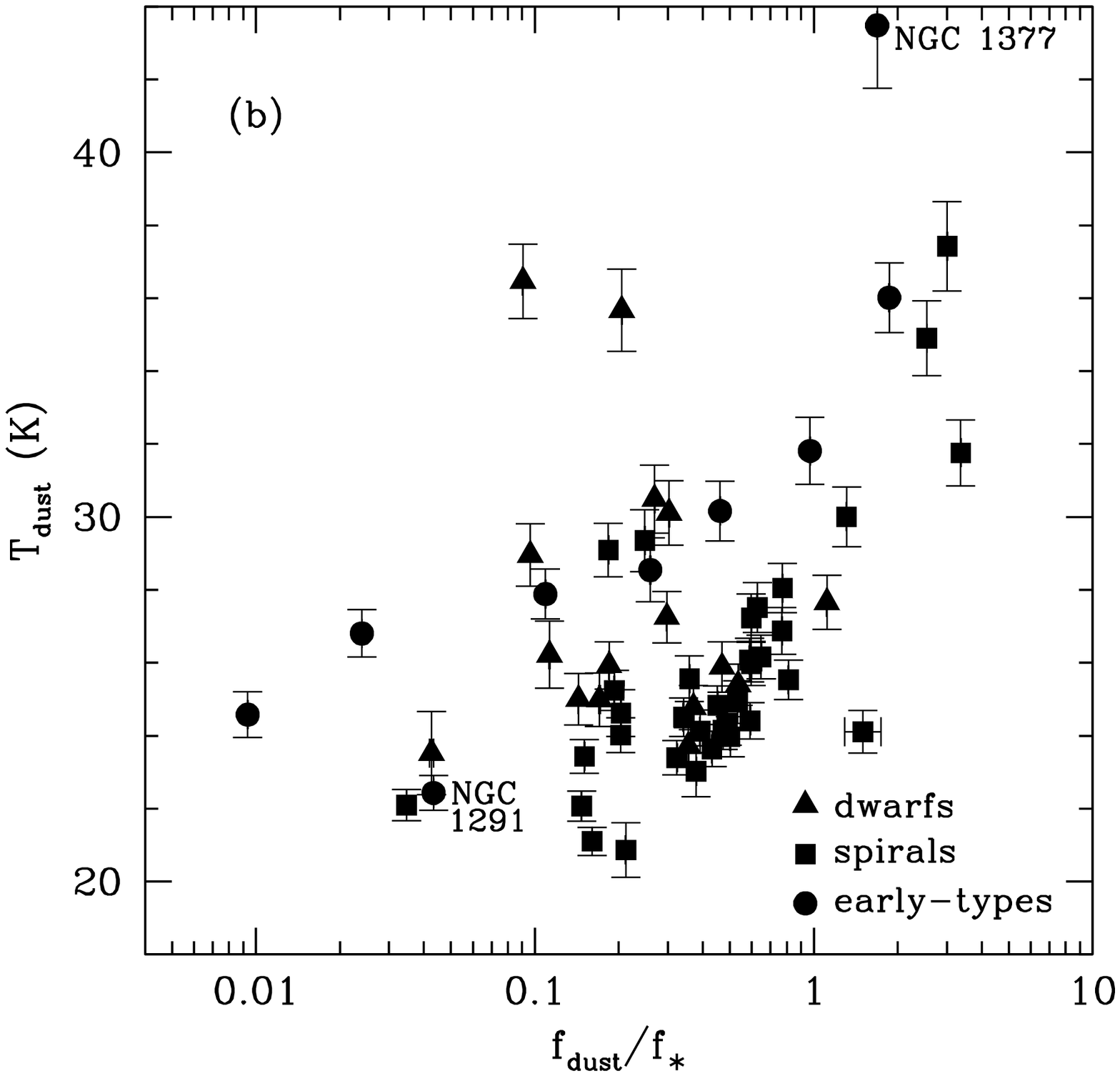} 
 \caption{
          Left plot shows MIPS $f_\nu(70\mu{\rm m})/f_\nu(160\mu{\rm m})$, 
          a proxy for dust temperature, versus dust/stellar flux ratio.  
          Right plot shows the dust temperature itself, estimated from a fit 
          to the far-IR SED with a single-temperature modified blackbody, vs. dust/stellar flux ratio. 
          Galaxies with E and S0, Sa to Scd, and Sd and later-type morphologies 
          are indicated by red circles, green squares, and blue triangles, respectively. 
          }
 \label{dusttemp}
\end{figure*}

We now examine the far-infrared colors of the galaxies in our sample, 
using MIPS and SPIRE bands.
`Warmer' far-IR colors are often associated with higher temperatures 
of small or large dust grain components, depending on the wavelengths (e.g., Li \& Draine 2001; 
Boselli et al.\ 2010b). 
Low $f_\nu(70\mu{\rm m})/f_\nu(160\mu{\rm m})$ color, for example, indicates 
that the far-IR SED peaks at a long wavelength, likely due to a large 
fraction of cold dust.

In Figure~\ref{dusttemp}, we show the correlation between the dust/stellar flux ratio and 
$f_\nu(70\mu{\rm m})/f_\nu(160\mu{\rm m})$.  We also show the correlation with dust 
temperature, described in Section~\ref{masses}, which uses these bands as well as the 
SPIRE flux densities.  The correlations between the dust/stellar flux ratio and SPIRE colors 
(not shown) have similar trends, but with more noise. 
As stated in Section~\ref{TdMd}, the temperatures depend on the assumed emissivity (e.g., Bendo et al.\ 2003), 
though the relative distribution of temperatures is robust. 
NGC~1377 is the galaxy with the warmest dust temperature, as mentioned in the previous section, 
as well as the largest $f_\nu(70\mu{\rm m})/f_\nu(160\mu{\rm m})$ color. 
The trends in both figures are similar because an increased $f_\nu(70\mu{\rm m})/f_\nu(160\mu{\rm m})$ color generally corresponds to a warmer dust temperature, even though we are simply fitting 
a single temperature modified blackbody.

In both figures, there is a correlation, but with considerable scatter (Spearman 
rank $r_s=0.33$ for the color plot and $r_s=0.36$ for the $T_\mathrm{dust}$ plot, with 95\% significance).  
In particular, only 
early-type galaxies ($r_s=0.68$ and $r_s=0.78$ for the two plots) and 
spirals ($r_s=0.66$ and $r_s=0.54$) exhibit a significant (3-$\sigma$) correlation: galaxies with relatively high dust emission also have relatively high far-IR colors and warm temperatures. 

In addition, at a given $f_\mathrm{dust}/f_\ast$ ratio, early-types have 
warmer colors and dust temperatures (by up to 5~K) than spirals. 
(The single exceptional early-type is NGC~1291, which Hinz et al.\ (in prep.) show 
has an outer ring that is dominated by much cooler dust than the rest of the galaxy.) 
Perhaps this occurs because the early-types in KINGFISH tend to have slightly smaller dust 
masses, which require less of a heating source (e.g., stars) to significantly raise their temperatures. 
It has long been known that dust grains may be heated by old stars (see, e.g., 
Helou 1986; Sauvage \& Thuan 1994; Calzetti et al.\ 1995; Kennicutt 1998; Draine \& Li 2001); 
however, the clearly different $T_\mathrm{dust}\,-\,f_\mathrm{dust}/f_\ast$ 
trend for early-types and spirals is a new result. 
The S0's with surprisingly large $f_\mathrm{dust}/f_\ast$ also 
have relatively warm dust temperatures, as well 
as slightly larger specific star formation rates, which we show in Section~\ref{SFRsec}. 
It appears that some S0's in the sample (as well as a couple Sa's) are undergoing 
a period of obscured star formation, and may be similar to the early-types examined 
by Shapiro et al.\ (2010) and Wei et al.\ (2010), which have molecular gas possibly accumulated 
through minor mergers, as well as slightly elevated star formation efficiencies.  
%
Nevertheless, most of the early-types with $f_\mathrm{dust}/f_\ast<1$ 
have very little ongoing star formation and lower stellar fluxes than 
their spiral counterparts, and yet they still have slightly warmer temperatures. 
Hence, these early-types may simply have more intense radiation fields. 
Early-types have many stars in their bulge components (as indicated, for example, by their 3.6 $\mu$m surface brightnesses), and these regions tend to have 
warmer dust temperatures (Engelbracht et al.\ 2010) in spite of the lower star formation rates, 
so \textit{we argue that a more intense interstellar radiation field is the most likely 
explanation of the different trends for late- and early-types.} 
This issue is currently being investigated further, using models of the dust masses and starlight intensities (Aniano et al., in prep.). 

Finally, the dwarf and irregular galaxies are scattered in the figures, although  
as noted by Walter et al.\ (2007) and Dale et al.\ (2007), some dwarfs
have high $f_\nu(70\mu{\rm m})/f_\nu(160\mu{\rm m})$ ratios,
indicating strong overall heating of the dust grain population.

\subsection{Star Formation Rate}\label{SFRsec}

\subsubsection{Estimating $SFR_{H\alpha+24\mu{\rm m}}$ and $SFR_{\mathrm{FUV}+\mathrm{TIR}}$}

Finally, we turn to the star formation rates (SFRs) of the galaxies in our sample. 
One way to estimate SFRs is to combine H$\alpha$ and mid-IR (specifically, 24 $\mu$m) 
luminosities (Calzetti et al.\ 2007; Zhu et al.\ 2008; Kennicutt et al.\ 2009).
We use the calibration proposed by Calzetti et al.\ (2010), 
with scatter $<\,0.2$ dex, for galaxies with a wide range of metallicities:
\begin{eqnarray}
  SFR\,(M_\odot/{\rm yr})&=&\,C_{{\rm H}\alpha}\,[L({\rm H}\alpha)_\mathrm{obs}+a_1 L(24)] \label{normalSFR} \\
   &\phantom{=}& \mathrm{if}\,\, L(24)<4\times 10^{42}\,{\rm erg}/{\rm s}, \nonumber\\
       &=&\,C_{{\rm H}\alpha}\,[L({\rm H}\alpha)_\mathrm{obs}+a_2 L(24)] \label{starburstSFR} \\
   &\phantom{=}& \mathrm{if}\,\, 4\times10^{42}\leq L(24) < 5\times10^{43}\,{\rm erg}/{\rm s} \nonumber
\end{eqnarray}
where the luminosities are in units of $\mathrm{erg}\,\mathrm{s}^{-1}$, 
$C_{{\rm H}\alpha}=5.45\times10^{-42}(M_\odot\,{\rm yr}^{-1})/(\mathrm{erg}\,\mathrm{s}^{-1})$, 
$a_1=0.020$, and $a_2=0.031$.
The first of these equations (\ref{normalSFR}) is calibrated for normal galaxies 
(Kennicutt et al.\ 2009), while the latter (\ref{starburstSFR}) is calibrated for HII 
regions and starbursts (Calzetti et al.\ 2007).  Based on the 24 $\mu$m luminosity, 
most of the KINGFISH galaxies (42/62, and 41/54 with H$\alpha$ fluxes) qualify as `normal' galaxies. 

For comparison, we also estimate SFRs using IR and UV luminosities, 
which is a complementary way to account for both obscured and unobscured star formation 
(e.g., Zheng et al.\ 2007; Buat et al.\ 2007). 
We use the SFR parameterization of Hao et al.\ (2011), which 
combines far-UV (1500~\AA) and TIR (Eqn.~\ref{DL07LTIR}) luminosities, 
and which updates the calibration in Dale et al.\ (2007): 
\begin{equation}
  SFR\,(M_\odot/{\rm yr}) \,=\, C_\mathrm{FUV}\, L_\mathrm{FUV}\,(1+a\,L_\mathrm{TIR}/L_\mathrm{FUV}), 
\end{equation}
where the luminosities are in units of $\mathrm{erg}\,\mathrm{s}^{-1}$, 
$C_\mathrm{FUV}=4.30\times10^{-44}(M_\odot\,{\rm yr}^{-1})/(\mathrm{erg}\,\mathrm{s}^{-1})$, and $a=0.475$.  
Like the stellar masses, we have assumed a Kroupa (2001) IMF for the SFRs. 

There is some dependence on the assumed star formation history (SFH), such as a SFH with a short recent starburst or with constant or declining star formation over a long time-scale, 
but the largest systematic uncertainty is the assumed (universal) IMF.  There also are other systematic uncertainties due to metallicity and AGN activity and to assumptions about the attenuation correction (see Kennicutt et al.\ 2009; Calzetti et al.\ 2010). 

Both star formation rates for the KINGFISH sample are listed in Table~\ref{table1}, 
except for the galaxies lacking $H\alpha$ or far-UV fluxes. 
The errors listed in the table are only the formal errors due to the flux densities; systematic uncertainties contribute at least 0.2 dex of additional uncertainty. 
The SFRs of the galaxies in the sample range from $10^{-3}$ to $10$ $M_\odot~\mathrm{yr}^{-1}$. 

The two SFR estimates are generally consistent, within 0.3 dex of each other. 
There are a few galaxies with larger discrepancies, such as NGC~1377 (although we had to perform an uncertain extrapolation for the far-UV flux), NGC~3351, NGC~4736, and NGC~6946 (although 
it has substantial diffuse background emission). 
IC~342, which is close to the Galactic plane ($l=138\degr$, $b=10\degr$), is dominated by obscured star formation, and the discrepancy between its SFRs appears to be due to a relatively low 24~$\mu$m flux. Consequently, its FUV$+$TIR SFR is likely more accurate. 
There are also discrepancies for low-mass dwarf irregular galaxies with very low SFRs (IC~2574, DDO~154, and DDO~165; see Walter et al.\ 2007), 
which are difficult to determine accurately. 


Our SFR and stellar mass estimates (discussed in Section~\ref{mstar}) 
can be compared to those obtained from model fits to the SEDs. 
%
For example, Johnson et al.\ (2007) used the stellar population models of Bruzual \& Charlot (2003) 
and the dust models of Witt \& Gordon (2000). 
Using similar models, we generally find excellent agreement for the stellar masses, and good agreement for the SFRs, which are consistent within 0.3 dex.  Two exceptions are NGC~3351, for which the SED fit yields a SFR that is 0.6 dex larger than our estimate using $H\alpha+24\mu$m luminosities, and NGC~3521, for which the fit yields a SFR that is 0.6 dex smaller. 

We also compare to da Cunha et al.\ (2008), who use 
the same stellar population models with exponentially declining star formation histories with bursts, and with the dust emission modeled as the sum of modified blackbodies at different temperatures.  
We also find excellent agreement for the stellar masses, but the SFRs of nine galaxies (out of 57) are statistically inconsistent and deviate by more than 0.4 dex. 
The most extreme case is NGC~3190, for which da Cunha et al.\ obtain $\mathrm{log}~\mathrm{SFR}=-1.23$, more than 0.8 dex lower than our estimates. 
%
Finally, Noll et al.\ (2009) also estimated stellar masses and SFRs of a sample of galaxies, 
32 of which are in our KINGFISH sample. 
They used the Maraston (2005) stellar population models, assuming either constant star formation or an exponential decreasing SFR.  Their SFRs are mostly consistent with ours, except for 
NGC~4536 and NGC~4736, for which they obtain values much larger than our SFR($H\alpha+24\mu$m) but similar to our SFR(TIR$+$FUV). 
We conclude that our SFRs are generally reliable, although for some galaxies they are 
difficult to determine accurately within a factor of two.


\subsubsection{Results}

\begin{figure*}
 \includegraphics[width=0.497\hsize]{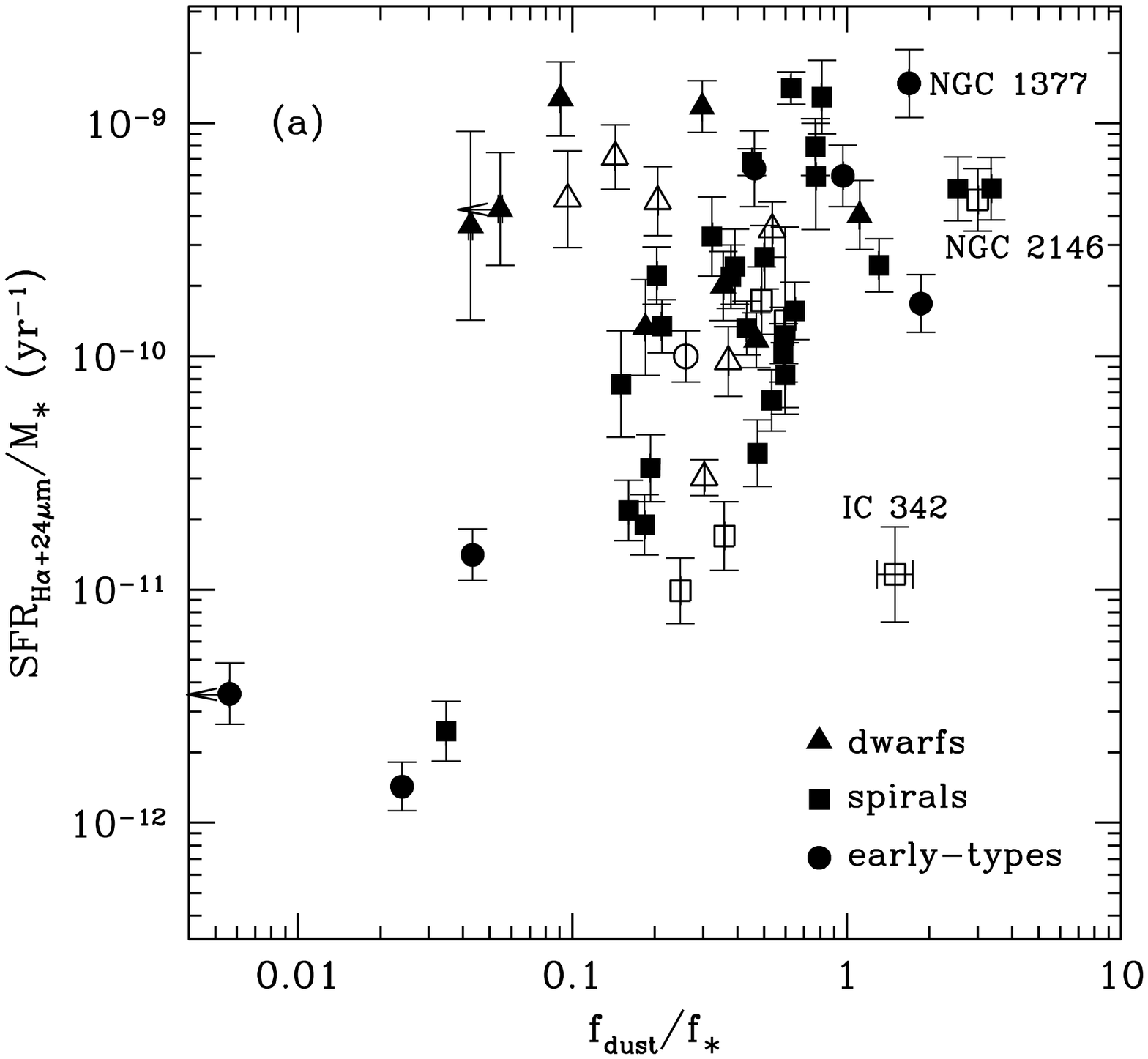} 
 \includegraphics[width=0.497\hsize]{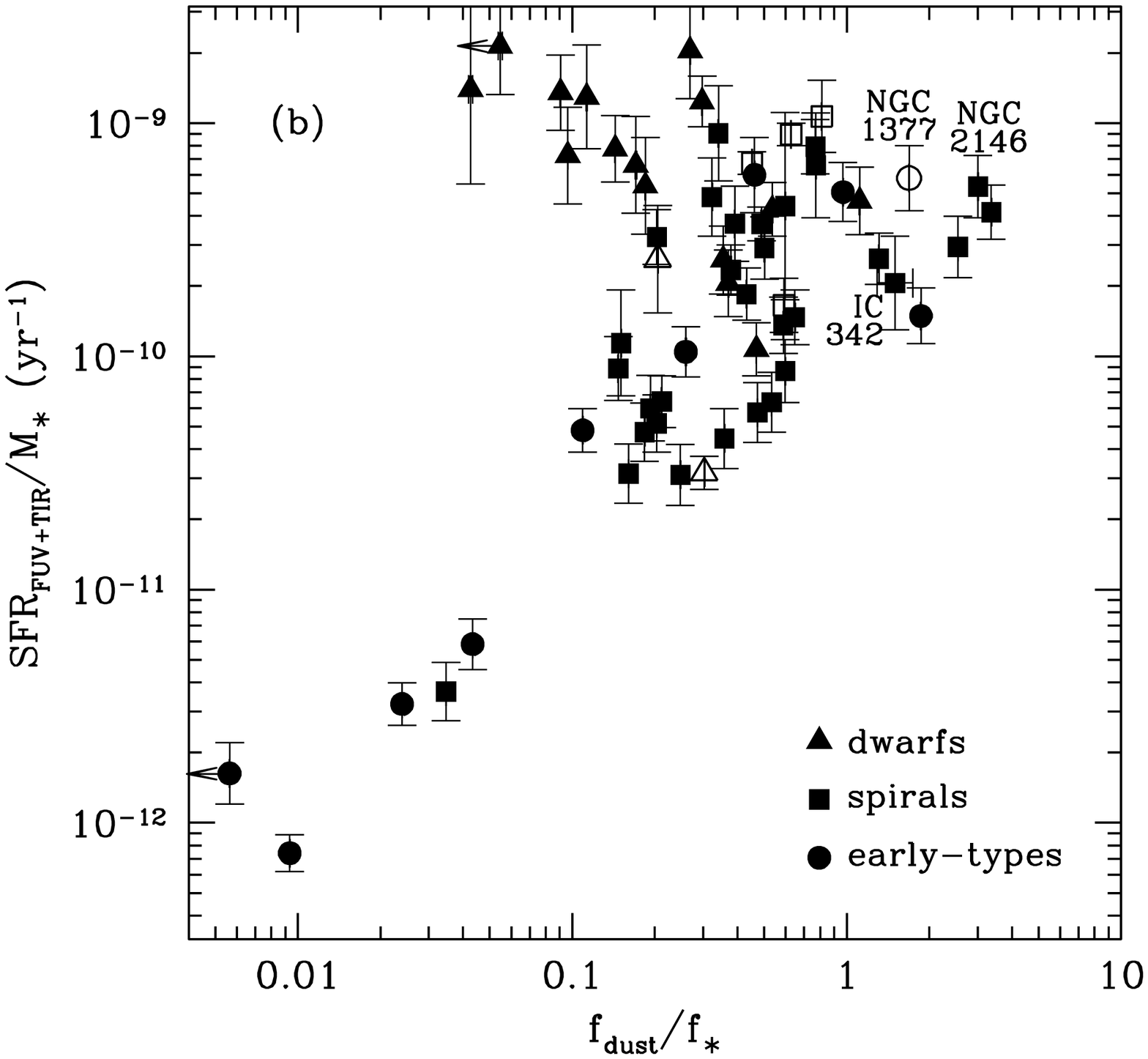} 
 \caption{Correlation between specific star formation rate and dust/stellar flux ratio. 
          The red circles, green squares, and blue triangles indicate E and S0, Sa to Scd, 
          and Sd and later-type morphologies, respectively.
          Left plot: SFR from combination of H$\alpha$ and 24 $\mu$m luminosities; open points
          indicate galaxies whose H$\alpha$ fluxes were obtained from Kennicutt et al.\ (2008).
          Right plot: SFR estimated from combination of TIR and FUV luminosities; open points 
          indicate galaxies whose FUV fluxes were extrapolated from longer wavelengths (see Section~\ref{dusttostellar}).  
         }
 \label{SSFRcorr}
\end{figure*}

We now show the correlation between specific SFR, or SFR per unit stellar mass, and 
dust/stellar flux ratio in Figure~\ref{SSFRcorr}, using both SFR estimators. 
The advantage of using SFR/$M_\ast$ over SFR is that it allows us to fairly compare 
galaxies with a wide range of stellar mass.

In general, there is a weak trend such that specific SFR increases with the 
dust/stellar flux ratio.  
In other words, galaxies with relatively bright dust emission also tend to have more star formation. 
The scatter appears to be larger using SFR$_{H\alpha+24\mu{\rm m}}$, although a 
statistical analysis indicates that its correlation is of similar strength 
(Spearman rank $r_s=0.43$, versus $0.30$ for SFR$_{{\rm FUV}+{\rm TIR}}$).  
Early-type and spiral galaxies exhibit particularly tight correlations 
($r_s\approx0.7$ and $r_s\approx0.65$, respectively, with $99\%$ significance). 
The scatter in the figure may be due to a number of factors, such as different 
contributions from obscured and unobscured tracers of star formation.

The correlation between the dust/stellar flux ratio and specific SFR is interesting in 
the context of its correlation with stellar mass and metallicity (see Figs.~\ref{fratiovsM}b 
and \ref{Mratios}b), considering that some authors have recently argued for a `fundamental 
plane' relating stellar mass, SFR, and metallicity of emission-line galaxies (Mannucci et al.\ 
2010; Lara-L\'{o}pez et al.\ 2010).  
Many of the KINGFISH galaxies would lie on this plane, but with substantial scatter and with 
outliers, among which would be the early-type galaxies lacking prominent emission lines.
These three galaxy properties---stellar mass, SFR, and metallicity---are certainly related 
to the dust/stellar flux ratio, 
but the inter-relations between star formation, stellar mass growth, 
dust production and metal enrichment are complex.


We can compare our results in Figure~\ref{SSFRcorr} to da Cunha et al.\ (2010), 
who studied similar properties of SDSS galaxies by modeling their SEDs.  
They similarly obtain a correlation between $M_\mathrm{dust}/M_\ast$ and SFR$/M_\ast$, 
although their correlation appears to be slightly stronger. 
They argue that stellar mass is not the main driver of this correlation. 
Nonetheless, we find that plotting the dust/stellar flux ratio against SFR 
(without normalizing by $M_\ast$) yields a shallower correlation with more scatter. 
da Cunha et al.\ (2010) also find that SFR$/M_\ast$ is strongly correlated with the dust-to-gas 
ratio and the fraction of $L_\mathrm{TIR}$ contributed by dust in the ambient ISM. 
In any case, our results appear to be consistent with theirs, with small 
differences likely due to sample selection (their sample is dominated by star-forming galaxies) 
and to the fact that their SEDs do not probe wavelengths longer than 100$\mu$m, which could 
yield underestimates of the dust fluxes and masses for some galaxies. 

The dwarf and irregular galaxies in Figure~\ref{SSFRcorr} depart from the positive correlation between specific SFR and dust/stellar flux ratio that we see for earlier-type galaxies 
($r_s=-0.33$ for SFR$_{H\alpha+24\mu{\rm m}}$ and $-0.70$ for SFR$_{{\rm FUV}+{\rm TIR}}$; 
the latter anti-correlation is statistically significant). 
Somewhat similarly, Dale et al.\ (2007) find that dwarf/irregular galaxies have specific
SFRs that decrease with the infrared-to-ultraviolet ratio (their Fig. 10). 
The dwarf galaxies' location in Figure~\ref{SSFRcorr} is most likely due 
to their small stellar masses, and their limited dust content (Walter et al.\ 2007); only 
a small fraction of dwarfs have large SFRs (Lee et al.\ 2009). 

Late-type and early-type galaxies have a similar relation between SFR/$M_\ast$ and 
$f_{\rm dust}/f_\ast$. 
Nonetheless, it is interesting that, according to Figure~\ref{dusttemp}, late-types 
have cooler dust temperatures (by up to 5~K) at a given dust/stellar flux ratio. 
For example, some S0 and Sa galaxies have similar specific SFRs and $f_\mathrm{dust}/f_\ast$, 
but the S0s have slightly warmer dust and lower dust masses.  
Perhaps in some of these galaxies, 
a small amount of star formation can more efficiently heat the dust; 
it is more likely, however, that 
the heating by the general ISRF, as well as by massive stars, is sufficient to heat the dust grains slightly more (e.g., see Draine \& Li 2007).

It is also interesting that approximately half of the early-type galaxies in the KINGFISH 
sample still have on-going star formation and dust heating, and are more similar to some of 
their late-type counterparts than to the passive early-types. 
This is contrary to the view that galaxies with large bulge components 
quickly `quench' their star formation.  
Fabello et al.\ (2011) argue against this view, showing that 
early-types with large bulge-to-disk ratios do not have lower atomic gas 
contents or star formation efficiencies than similar later type galaxies. 
Crocker et al.\ (2011) account for both atomic and molecular gas and obtain a similar result, 
such that star-forming early-type galaxies lie in a 
similar range of the Schmidt-Kennicutt relation as normal star-forming galaxies. 
Bulge components may be a necessary condition for suppressing star formation, but 
they are not sufficient (Bell 2008). 

A number of KINGFISH galaxies appear to be similar to `passive disks' and 
`red spirals'\footnote{`Red spirals' are spiral galaxies on the red sequence, as defined by optical colors.  They include both passive spirals as well as spirals with obscured star formation (Masters et al.\ 2010b).} 
recently studied by other authors (e.g., Wolf et al.\ 2009; Skibba et al.\ 2009; van der 
Wel et al.\ 2009; Masters et al.\ 2010b; Bundy et al.\ 2010). 
Ten galaxies in the sample have dust masses greater than $10^8\,M_\odot$, 
and all of them are spirals.  The majority (7/10) of the galaxies with 
$f_\mathrm{dust}/f_\ast>0.75$ are spirals as well. 
Of the spiral galaxies with large dust masses or large dust/stellar ratios, 
some have low specific SFRs, such as the earlier-type spirals NGC~4725 and NGC~1512. 
These may be examples of `transition' galaxies discussed by Masters et al.\ (2010b) and Bundy et al.\ (2010), galaxies between star-forming disk-dominated galaxies and passive early-types.  

Nonetheless, large dust mass is not a sufficient condition for declining SFRs, because there 
are dusty star-forming galaxies such as the late-types NGC~4254 and NGC~4631. 
Earlier-type spirals with obscured star formation, such as NGC~1097, NGC~2146, NGC~2798, 
and NGC~7331, have also been described as transition galaxies (Zhu et al.\ 2011). 
In addition, a galaxy need not have a dominant bulge to have passive star formation: 
NGC~5055 and NGC~4321 are dusty galaxies with very low specific SFRs, but with 
Sbc morphologies. 
%
Finally, the KINGFISH sample also contains 
massive galaxies with `pseudobulges' (i.e., lacking classical bulges), 
like NGC~5457 and NGC~6946, which are difficult to explain with current models of galaxy formation, in which most massive galaxies experience major mergers (Kormendy et al.\ 2010); 
however, they both have a very large gas supply (Walter et al.\ 2008) with which they could 
potentially form stars, and 
NGC~6946 may also have a particularly high star formation efficiency in its spiral arms (Foyle et al.\ 2010). 
It appears that, while some galaxies may be undergoing a transition to passive star formation 
and earlier-type morphologies, the transition is blurry and some galaxies evolve differently 
than others.


%
%

%

\section{Extragalactic Background Light}\label{EBL}


Our dust/stellar flux ratios quantify the relative contributions of emission from dust 
and stars of individual galaxies.  The extragalactic background light (EBL) is a 
related concept, because it can be used to quantify the contributions of stellar- 
and dust-dominated galaxies to the optical and infrared background.  Rather than 
examining SEDs within galaxies, analyses of the EBL involve integrating the 
light from many background galaxies. 
Studies of the EBL distinguish between optical light from stars and IR light from dust, 
so it is useful to compare these studies to our analysis of the stellar and dust emission of individual galaxies, 
although because of the small size and incompleteness of the KINGFISH sample, we 
cannot make strong conclusions based on these comparisons.

A number of authors have investigated the EBL and have quantified it at different wavelengths
(e.g., Hauser et al.\ 1998; Hauser \& Dwek 2001; Dole et al.\ 2006; B\'{e}thermin et al.\ 2010; Kneiske \& Dole 2010). 
A recent study (Dom\'{i}nguez et al.\ 2011) has attempted to distinguish between the contributions 
of different types of galaxies to the EBL. 
There has also been some theoretical work on explaining the intensity of the EBL, in the 
context of the evolving stellar mass density and SFR density 
(e.g., Chary \& Elbaz 2001; Primack et al.\ 2005; Nagamine et al.\ 2006; Fardal et al.\ 2007). 
For a recent comparison of some models and constraints on the IR EBL, see Orr et al.\ (2011).
%
The EBL is the dominant radiant energy in the universe aside from the cosmic 
microwave background, and it is thought to be mostly due to stars, dust, and AGN 
in galaxies (e.g., Hauser \& Dwek 2001), while faint diffuse emissions can represent 
only a small fraction of the integrated energy (Dole et al.\ 2006).

Some have attempted to compare the cosmic infrared background (CIB) and the cosmic optical background (COB). Hauser \& Dwek (2001) found that $\approx52\%$ of the EBL is contributed 
by direct starlight (and the rest absorbed and re-emitted by dust at $\lambda\geq3.5\mu$m), but with large uncertainties, 
while Dole et al.\ (2006) found that the ratio COB/CIB ranges from 0.7 to 1.5 (making the demarcation at $\lambda=8\mu$m). 
Perhaps not coincidentally, some of the galaxies in the KINGFISH sample have similar 
ratios of $f_\mathrm{dust}/f_\ast$, which is analogous to CIB/COB. 

The galaxies that dominate the COB and CIB may constitute different populations; for example, 
Dole et al.\ (2006) argue that the CIB is mainly composed of luminous infrared galaxies (LIRGs) 
at $z\sim1$. 
Galaxies in the KINGFISH sample with bright TIR luminosity, large masses and specific SFRs, 
such as NGC~1482, NGC~2146, NGC~2798, and NGC~7331, could be considered to be 
examples of low-redshift counterparts of 
these galaxies, although they are not as luminous as LIRGs. 
KINGFISH also includes galaxies which may be counterparts of objects dominating the 
optical EBL, such as NGC~3521 and NGC~5055, which are 
massive galaxies with significant star formation. 

Lastly, note that the mean stellar/dust flux ratio of the galaxies in our sample is 
$\overline{f_\mathrm{dust}/f_\ast}=0.55$, and split by morphological  
type, the means are 0.28, 0.69, and 0.54 for dwarfs, spirals, and early-types, respectively. 
The summed $f_\mathrm{dust}/f_\ast$ ratio can be more directly compared to the CIB/COB estimates 
of Hauser \& Dwek (2001) and Dole et al.\ (2006).  
This quantity, which is dominated by the more luminous galaxies, is 
$\Sigma(f_\mathrm{dust})/\Sigma(f_\ast)=0.44\pm0.01$ for the whole sample. 
This can also be expressed as 
$\Sigma(f_\mathrm{dust})/\Sigma(f_\ast+f_\mathrm{dust})=0.30\pm0.01$. 
Split by type, this dust fraction is 0.34, 0.32, and 0.07 for dwarfs, spirals, and 
early-types, respectively. 
KINGFISH does not contain a statistically complete sample; in general, dwarf galaxies 
dominate by number, although they are typically faint (e.g., Lee et al.\ 2009).  
Many early-types are luminous and stellar-dominated, but they are relatively few in 
number (e.g., Dale et al.\ 2009). 
Our results appear to indicate that, at least for nearby galaxies, spirals may constitute the most important contribution, for which their stars contribute $\approx68\%$ of the light and their dust contribute $\approx32\%$. 
This is approximately consistent with Soifer \& Neugebauer (1991), who analyzed a 
flux density-limited sample of local galaxies with IRAS and obtained a total dust/stellar fraction  
of $\approx23\%$, which is smaller than our result, probably due to their limited coverage in the far-IR.  
In any case, the local fraction is smaller than the high-redshift infrared EBL fraction estimated by Dole et al.\ 
(2006; CIB/(COB+CIB)$\approx40-60\%$), which indicates that 
the IR output of galaxies evolves with time and was larger in the past, consistent with studies 
of the evolution of IR luminosity functions (e.g., Le Floc'h et al.\ 2005; Chary \& Pope 2010; Murphy et al.\ 2011a; cf., models in Fontanot \& Somerville 2010). 
Nevertheless, a larger and more complete low-redshift catalog would be required to investigate this issue further.

\section{Summary}\label{discussion}

We now summarize our main conclusions:

\begin{itemize}



\item The dust/stellar flux ratios estimated from global galaxy SEDs are correlated with total-infrared luminosity, in a morphology and metallicity dependent way. 
Metal-poor dwarf galaxies tend to have faint IR luminosities, while spirals tend to have 
lower metallicities and higher dust/stellar flux ratios than early-types.

\item Dust/stellar flux and dust/stellar mass ratios are correlated, especially for early-types. Late-types and dwarf galaxies show considerable scatter, partly due to the effect of metallicity on the flux ratio.  Some of the scatter is also due to the dependence on dust temperature.

\item Most galaxies exhibit a trend such that, those with large dust/stellar flux ratios 
have warmer FIR colors and dust temperatures. Late-types tend to have slightly cooler 
temperatures (by up to 5~K) than early-types at a given dust/stellar flux ratio, while dwarf and irregular galaxies have more scattered temperatures. 

\item We find that late- and early-type galaxies have specific SFRs 
that are correlated with dust/stellar flux ratios: galaxies with more dust 
emission also tend to have more star formation. 
Combined with our previous result, we interpret this as evidence that ongoing star 
formation is sufficient to heat some of the dust in these galaxies, while other galaxies 
have more intense radiation fields, where the older stellar population likely 
contributes significantly to the dust heating.

\item The KINGFISH sample contains a number of dusty star-forming spiral galaxies as well 
as some passive spirals, whose 
limited star formation resembles that of some early-types. 

\end{itemize}
Finally, we note that our results could contribute constraints for galaxy formation models, such as 
on the amount of dust production, metal enrichment, and star formation in different types 
of low-redshift galaxies. 
In addition, our results could be useful as a local benchmark for comparisons with high-redshift studies, 
such as studies of submillimeter galaxies (e.g., Santini et al.\ 2010) and of the evolution of the dust and stellar content of galaxies (e.g., Dunne et al.\ 2011). 

\vspace{0.2cm}
\section*{Acknowledgments}

We thank John Moustakas for providing and discussing the oxygen abundances 
of Moustakas et al.\ (2010). 
We thank Andy Marble for valuable discussions about our results. 
We also thank the anonymous referee for insightful comments that helped to improve the quality of the paper.


Herschel is an ESA space observatory with science instruments provided by European-led Principal Investigator consortia and with important participation from NASA. SPIRE has been developed by a consortium of institutes led by Cardiff University (UK) and including Univ. Lethbridge (Canada); NAOC (China); CEA, LAM (France); IFSI, Univ. Padua (Italy); IAC (Spain); Stockholm Observatory (Sweden); Imperial College London, RAL, UCL-MSSL, UKATC, Univ. Sussex (UK); and Caltech, JPL, NHSC, Univ. Colorado (USA). This development has been supported by national funding agencies: CSA (Canada); NAOC (China); CEA, CNES, CNRS (France); ASI (Italy); MCINN (Spain); SNSB (Sweden); STFC (UK); and NASA (USA).




\end{document}